\documentclass[11pt,preprintnumbers,nofootinbib,amsmath,amssymb]{revtex4}
\usepackage{float}
\usepackage{indentfirst}
\usepackage{amsmath,amssymb,epsfig,subfigure}
\usepackage{graphicx}
\usepackage[export]{adjustbox}
\usepackage{color}
\usepackage{multirow}
\usepackage{booktabs}
\usepackage{changes}
\usepackage{footnote}
\usepackage{mathrsfs} 
\usepackage[shortcuts]{extdash}

\usepackage{hyperref}
\hypersetup{colorlinks=true,linkcolor=blue,citecolor=magenta}

\begin{document}
\renewcommand{\baselinestretch}{1.15}
\makeatletter
\renewcommand{\thesubfigure}{\alph{subfigure}}
\renewcommand{\@thesubfigure}{(\thesubfigure)\hskip\subfiglabelskip}
\makeatother

\title{Scalarization of Bardeen spacetime}

\preprint{}

\author{Long-Xing Huang and Yong-Qiang Wang\footnote{E-mail: yqwang@lzu.edu.cn, corresponding author}}

\affiliation{$^{1}$ Lanzhou Center for Theoretical Physics, Key Laboratory of Theoretical Physics of Gansu Province, School of Physical Science and Technology, Lanzhou University, Lanzhou 730000, China\\
 $^{2}$ Institute of Theoretical Physics $\&$ Research Center of Gravitation, Lanzhou University, Lanzhou 730000, China}


\begin{abstract}

 In this paper, we study the scalarization of the entire Bardeen spacetime which is constructed from a nonlinear magnetic monopole. We find that once the scalarization coupling parameter exceeds the scalarized threshold $a_t$, a scalarized Bardeen spacetime (SBS) solution exists for any magnetic charge $q$. However, the nature of the scalarization depends on the magnetic charge $q$. For $q$ less than a critical vaule $q_c$, when $a$ exceeds $a_t$, the scalar field emerges. As $a$ approaches $a_t$ from above (i.e., $a\to a_t^+$), the scalar field vanishes and the metric is reduced to a pure Bardeen spacetime. This behavior indicates that the solution exhibits the general ``spontaneous" scalarization phenomenon. Conversely, for $q \geq q_c$, in the limit $a \to a_t^+$, the scalar field is non-vanishing and the SBS approaches a ``frozen" SBS. Considering the importance of photon orbits in astronomical observations, we analyze the trajectories of photons around SBSs by analyzing null geodesics.
 
\end{abstract}


\maketitle
\newpage

\section{INTRODUCTION}\label{sec: introduction}
    Since its formulation more than 100 years ago, general relativity has achieved remarkable success in both theory and experiment. Black holes (BHs) are one of the most important predictions of general relativity and garnering significant attention. According to the no-hair theorem~\cite{Israel:1967wq,Carter:1971zc,Ruffini:1971bza}, for any object collapsing into a BH, regardless of its initial complexity, its final stable state is characterized by only three observable parameters: the mass $M$, angular momentum $J$, and electric charge $Q$. This means that all other information about the matter that formed the BH (called ``hair") is lost, making all black holes with the same ($M$, $J$, $Q$) indistinguishable. However, ``no-hair theorem" can be violated under certain conditions. For example, in 1988, the discovery of one type of asymptotically flat self-gravitating Yang-Mills soliton, shown by the pioneering work of R. Bartnik and J. Mckinnon~\cite{Bartnik:1988am}, which inspired the construction of various hairy BH solutions (see also the review~\cite{Herdeiro:2015waa,Volkov:2016ehx}). 
    
    Among the dynamical mechanisms leading to hairy BHs, the spontaneous scalarization has drawn widespread attention in recent years. In this process, when a certain parameter exceeds a certain threshold, the scalar field switches from a trivial constant configuration to a nontrivial one, which is analogous to the spontaneous magnetization of ferromagnets~\cite{Damour:1996ke}. One possible realization of this mechanism was first discovered by T. Damour and G. Esposito-Farese~\cite{Damour:1993hw} in the background of neutron stars in 1993. In their model, the scalar field couples to the matter field (i.e., the ``source term") through a coupling function $A(\varphi)$. In the spirit of Brans–Dicke theory~\cite{Will:1981,Salmona:1967zz,Hillebrandt:1974,saenz1977maximum}, such a nonminimal coupling can be viewed as a field-dependent gravitational constant~\cite{Salgado:1998sg}. When the stellar compactness exceeds a critical threshold, a “tachyonic instability” is triggered, leading to rapid scalar field growth and the generation of a scalar charge, until nonlinear effects quench the instability, and the outcome is a neutron star with nontrivial ``scalar hair"~\cite{Novak:1997hw}. 

    The study of spontaneous scalarization of BH began in the scalar–tensor Gauss–Bonnet theory~\cite{Doneva:2017bvd,Silva:2017uqg,Antoniou:2017acq,Minamitsuji:2018xde,Silva:2018qhn,Doneva:2019vuh,Macedo:2019sem,Antoniou:2021zoy}, where the Gauss–Bonnet invariant is coupled to the scalar field and serves as the source term. In fact, this mechanism is quite general. For instance, they can be extended to asymptotically AdS/dS BHs with a cosmological constant~\cite{Bakopoulos:2018nui,Brihaye:2019gla,Bakopoulos:2019tvc,Bakopoulos:2020dfg,Lin:2020asf,Guo:2020sdu}, rotating BHs~\cite{Cunha:2019dwb, Collodel:2019kkx} or spin-induced scalarization~\cite{Dima:2020yac,Doneva:2020nbb,Doneva:2020kfv,Herdeiro:2020wei}. Moreover, the source term can also be replaced with the other invariants, such as the Maxwell invariant~\cite{Herdeiro:2018wub}, the Ricci scalar~\cite{Herdeiro:2019yjy}, or the Chern-Simons invariant~\cite{Brihaye:2018bgc}, etc. Currently, a substantial body of literature has emerged on this aspect of research, which is attracting increasing attention and continues to make progress - see review~\cite{Doneva:2022ewd} and the literature therein.

    However, despite the recent observations of gravitational waves from the binary black-hole mergers~\cite{LIGOScientific:2016aoc} and direct imaging of black-hole shadows~\cite{EventHorizonTelescope:2019dse, EventHorizonTelescope:2019uob,EventHorizonTelescope:2019jan,EventHorizonTelescope:2019ths,EventHorizonTelescope:2019pgp,EventHorizonTelescope:2019ggy} by the Event Horizon Telescope (EHT), BHs in general relativity still suffer from a fatal drawback — the presence of spacetime singularities. The means that matter can be infinitely compressed, and all physical laws will break down here, which is unacceptable for a physical entity. According to the ``singularity theorem" proposed by Hawking and Penrose, singularities seem inevitable for a BH solution if matter obeys certain prerequisites such as the strong energy condition~\cite{Penrose:1964wq,Hawking:1970zqf}. However, this means introducing a suitable matter field that does not satisfy the strong energy condition may eliminate singularities. The first static, spherically symmetric regular BH with an unknown physical source violating the strong energy condition, later referred to the Bardeen BH, was proposed by J. Bardeen in 1968~\cite{Bardeen:1968} (see Ref.~\cite{Shirokov1948,Duan:1954bms,Sakharov:1966aja,Gliner1966} for early attempts about regular BHs and see Ref.~\cite{Ansoldi:2008jw,Lan:2023cvz,Torres:2022twv,Carballo-Rubio:2025fnc} for a review). More than three decades later, E. Ayon-Beato and A. Garcia reinterpreted the Bardeen model as the gravitational field of a nonlinear magnetic monopole, and hence identified the matter source of Bardeen BH as a nonlinear electromagnetic field, where the magnetic charge $q$ exceeds a certain critical value $q_c$ ~\cite{Ayon-Beato:1998hmi,Ayon-Beato:2000mjt}.

    Recently, the spontaneous scalarization of the Bardeen black hole was investigated in Ref.~\cite{Zhang:2024bfu}, which was restricted to solving the field equations in the region outside the horizon, without inside. However, the ``regularity” of a regular black hole is understood in a global sense, namely that the metric must remain finite throughout the entire spacetime. Therefore, this naturally raises a question: can scalarization occur throughout the entire spacetime? In this paper, we extend to study the model of the spontaneous scalarization of the Bardeen spacetime with various magnetic charges over the entire space region. In this context, we find that 
    once the scalarization coupling parameter exceeds the scalarized threshold $a_t$, a scalarized Bardeen spacetime (SBS) exists for any charge. In particular, when $q$ is smaller than the critical value $q_c$ and with $a\to a_t^+$, the scalar field vanishes and the solution exhibits the general “spontaneous” scalarization phenomenon. However, when the magnetic charge $q \geq q_c$, a special scalarized Bardeen spacetime solution can emerge with $a\to a_t^+$. For such solutions, there exists a special radius $r_{cH}$ within which the real scalar field is predominantly localized. For the metric field, $-g_{tt}$ approaches zero for $r<r_{cH}$, while $1/g_{rr}=g^{rr}$ becomes nearly zero at $r_{cH}$ (in fact, this phenomenon can also be observed in many other situations, see e.g.~\cite{Wang:2023tdz,Huang:2023fnt, Yue:2023sep,Ma:2024olw,Huang:2024rbg,Chen:2024bfj, Zhang:2024ljd,Wang:2024ehd,Sun:2024mke,Zhao:2025hdg,Huang:2025css,Zhang:2025nem,Brihaye:2025dlq,Zhao:2025yhy,Chicaiza-Medina:2025wul}). These properties of the metric imply that, from the perspective of an observer at infinity, objects near
    the critical horizon move exceedingly slowly, as if ``frozen.”  Hence, we refer to this frozen state as the ``frozen SBS". Finally, considering the importance of photon orbits in observations, we also analyze the photon orbits in SBSs by analyzing the null geodesics. 

    This paper is organized as follows. In Sect.~\ref{sec: model}, we present the general framework of the model SBSs. In Sect.~\ref{sec: bound}, we discuss the boundary conditions. The numerical results are provided in Sect.~\ref{sec: results}. Finally, the brief conclusion and discussion are given in Sect.~\ref{sec: conclusion}.

\section{THE GENERAL FRAMEWOEK}\label{sec: model}

In this section, we will give a brief introduction to the SBSs model, where the Bardeen nonlinear electromagnetic field is non-minimally coupled to a real, massless scalar field $\Phi$ through the coupling function $f(\Phi)$. The model has the following action to describe (we use units with $\hbar = c = 4\pi G = 1$):
    \begin{equation}
        I=\int d^4 x \sqrt{-g}\left[\frac{R}{4}-2\nabla_{\mu}\Phi\nabla^{\mu}\Phi+ f(\Phi) \mathcal{L}(\mathcal{F})\right],
        \label{eq:action}
    \end{equation}	
with 
		\begin{equation}
		    			\mathcal{L}(\mathcal{F})=-\frac{3}{2s}(\frac{\sqrt{2q^2\mathcal{F}}}{1+\sqrt{2q^2\mathcal{F}}})^{5/2}.  
         			\label{eq:lagrangianb}
		\end{equation} 
Here, $R$ represents the Ricci scalar. The electromagnetic field Lagrangian density $\mathcal{L}(\mathcal{F})$ is a function of $\mathcal{F}=\frac{1}{4}F_{\mu\nu}F^{\mu\nu}$ with the nonlinear electromagnetic field strength $F=\partial_{\mu}A_{\nu}-\partial_{\mu}A_{\nu}$, where $A_{\mu}$ is the Bardeen electromagnetic field. $s$ and $q$ are free parameters, where $q$ represents the magnetic charge. The field equations, obtained by varying the action (\ref{eq:action}) with respect to the metric $g_{\mu\nu}$, Bardeen electromagnetic field, and real massless scalar field, read as
\begin{equation}
    R_{\mu \nu} - \frac{1}{2}g_{\mu \nu}R - 2 T_{\mu \nu}=0,
    			\label{eq:einstein}
\end{equation}
\begin{equation}
    \nabla_{\mu}\left(f(\Phi)\frac{\partial \mathcal{L}}{\partial \mathcal{F}} F^{\mu \nu}\right)=0,
    			\label{eq:equationBardeen}
\end{equation}
\begin{equation}
    \nabla^2\phi=\frac{1}{4}\frac{\partial f(\Phi)}{\partial \Phi} \mathcal{
    L(\mathcal{F})
    },
    			\label{eq:equationKG}
\end{equation}
with the energy–momentum tensor
\begin{equation}
    T_{\mu \nu}=\frac{1}{2}f(\Phi)\left[-\frac{\partial \mathcal{L}}{\partial \mathcal{F}} F_{\mu\alpha}F^{\alpha}_{\nu}+g_{\mu\nu} \mathcal{L}(\mathcal{F})\right]+2\partial_{\mu}\Phi\partial_{\nu}\Phi-\nabla_{\mu}\Phi\nabla_{\nu}\Phi.
    \label{eq:emtensor}
\end{equation}
In this paper, we choose $f(\Phi)=e^{-\alpha \Phi^2}$. Restricting to the static, spherically symmetric solutions, we adopt the following metric ansatz 
\begin{equation}
    d s^2=-n(r) \sigma^2(r) d t^2+\frac{d r^2}{n(r)}+r^2\left(d \theta^2+\sin ^2 \theta d \varphi^2\right),
    \label{eq:metric}
\end{equation}
where the meteic functions $n(r)$ and $o(r)$ are only depend on the radial variable $r$. Furthermore, for the real scalar field and the electromagnetic field, we choose
\begin{equation}
    \Phi=\phi(r),\quad A_{\mu}dx^{\mu}=p\cos(\theta)d\varphi.
        \label{eq:field}
\end{equation}
Here, the function $\phi(r)$ is a radial real function. Substituting the above ansatzes (\ref{eq:metric}) and (\ref{eq:field}) into the equations of motion (\ref{eq:einstein}-\ref{eq:equationKG}), we can obtain three ordinary differential equations (ODE) for $n(r)$, $o(r)$ and $\phi(r)$:
\begin{align}
&n'+n\left(4 r\phi'^2+\frac{1}{r}\right)+\frac{3 e^{-a\phi^2}q^5r}{(q^2+r^2)^{5/2}s}-\frac{1}{r}=0\\
&o'-4 r o\phi'^2=0\\
&\phi''+\phi'\left(\frac{n'}{n}+\frac{o'}{o}+\frac{2}{r}\right)+\phi\frac{3ae^{-a\phi^2q^5}}{4ns(q^2+r^2)^{5/2}}
    \label{eq:finaleq}
\end{align}
It is worth noting that when $\phi=0$, the solution of Eq.~(\ref{eq:finaleq}) can be degenerates to the Bardeen solution, and the metric taking the following form:
\begin{equation}
d s^2=-g(r) d t^2+g(r)^{-1} d r^2+r^2\left(d \theta^2+\sin ^2(\theta) d \varphi^2\right),
\end{equation}
here, 
\begin{equation}
    f(r)=1-\frac{q^3r^2}{s(r^2+q^2)^{3/2}}.
    \label{eq:BardeenSolution}
\end{equation}
Here $M = q^3/2s$ is the mass~\cite{Ayon-Beato:1998hmi}. From Eq.~(\ref{eq:BardeenSolution}), one can find the function $g(r)$ exhibits a local minimum value at $ r=\sqrt{2}q$. In case where $q<q_c=3^{3/4}\sqrt{\frac{s}{2}}$ (For our setting where $s=0.2$, $q_c\approx0.7208$), the solutions without event horizons are present. When $q=q_c$, degenerate horizons are observed, and for $q>q_c$, two distinct horizons exist. Therefore, When $q \geq q_c$, the metric (\ref{eq:BardeenSolution}) represents the BHs, which is different from the case $q < q_c$.

\section{BOUNDARY CONDITION AND NUMERICAL IMPLEMENTATION}
\label{sec: bound}
To solve the system of ODEs (\ref{eq:finaleq}), appropriate boundary conditions, can be derived from assumptions of regularity and asymptotic flatness of the solution, must be established for each unknown function. For the metric functions, $n(r)$ and $\sigma(r)$ satisfy: 
\begin{equation}
    n(0)=1,\quad \sigma(0)=\sigma_0,\quad n(\infty)=1-\frac{2GM}{r},\quad \sigma(\infty)=1,
\end{equation}
where the value of $\sigma_0$ and the ADM mass $M$ of the solution are currently unknown, their can be determined by solving the ODEs system. In addition, for the scalar field, we require
\begin{equation}
    \left.\frac{d \phi}{d r}\right|_{r=0}=0,\quad \phi(\infty)=0.
\end{equation}
    From the asymptotic behavior of the scalar field, the so called ``scalar charge”\cite{Damour:1993hw} $Q_s$ can be defined as:
    \begin{equation}
        Q_s :=-\lim\limits_{r \to \infty} r^2 \frac{d\phi}{dr}
    \end{equation}

\section{NUMERICAL RESULTS}\label{sec: results}

To solve numerically performed over the entire spacetime region, we introduce a new radial coordinate
\begin{equation}
    \bar{r}=\frac{r}{r+1},
\end{equation}
which maps the radial coordinate range from the semi-infinite region $\left[0,\infty\right)$ to the unit interval $\left[0, 1\right]$. After obtaining the numerical solution, the inverse transformation $r=\frac{\bar{r}}{1-\bar{r}}$ can be used to replace the $\bar{r}$ coordinates with the $r$ coordinates. 

Furthermore, in order to facilitate numerical computations, without loss of generality, we set $s=0.2$. Then, the solution is controlled only by the magnetic charge $q$ and the parameter $a$.

We numerically solve the system of ODEs.~(\ref{eq:finaleq}) by employing the finite element method with $1000$ grid points distributed over the integration region $[0, 1]$. The iterative method we employ is the Newton-Raphson method, and to ensure the accuracy of the computed results, the relative error is required to be below $10^{-5}$.

In our numerical results, we find that SBSs will appear when the coupling parameter $a$ exceeds the thresholds $a_{t}$ (referred to as the ``scalarization thresholds"). In the case of $a<a_{t}$, the scalar field will vanish, and the solution of SBSs reduces to the pure Bardeen magnetic monopole theory. 

\subsection{ADM matter and scalar charge}\label{sec:Mw}

\begin{figure}[!htbp]
		\begin{center}
		\subfigure{ 
			\includegraphics[height=.28\textheight,width=.32\textheight, angle =0]{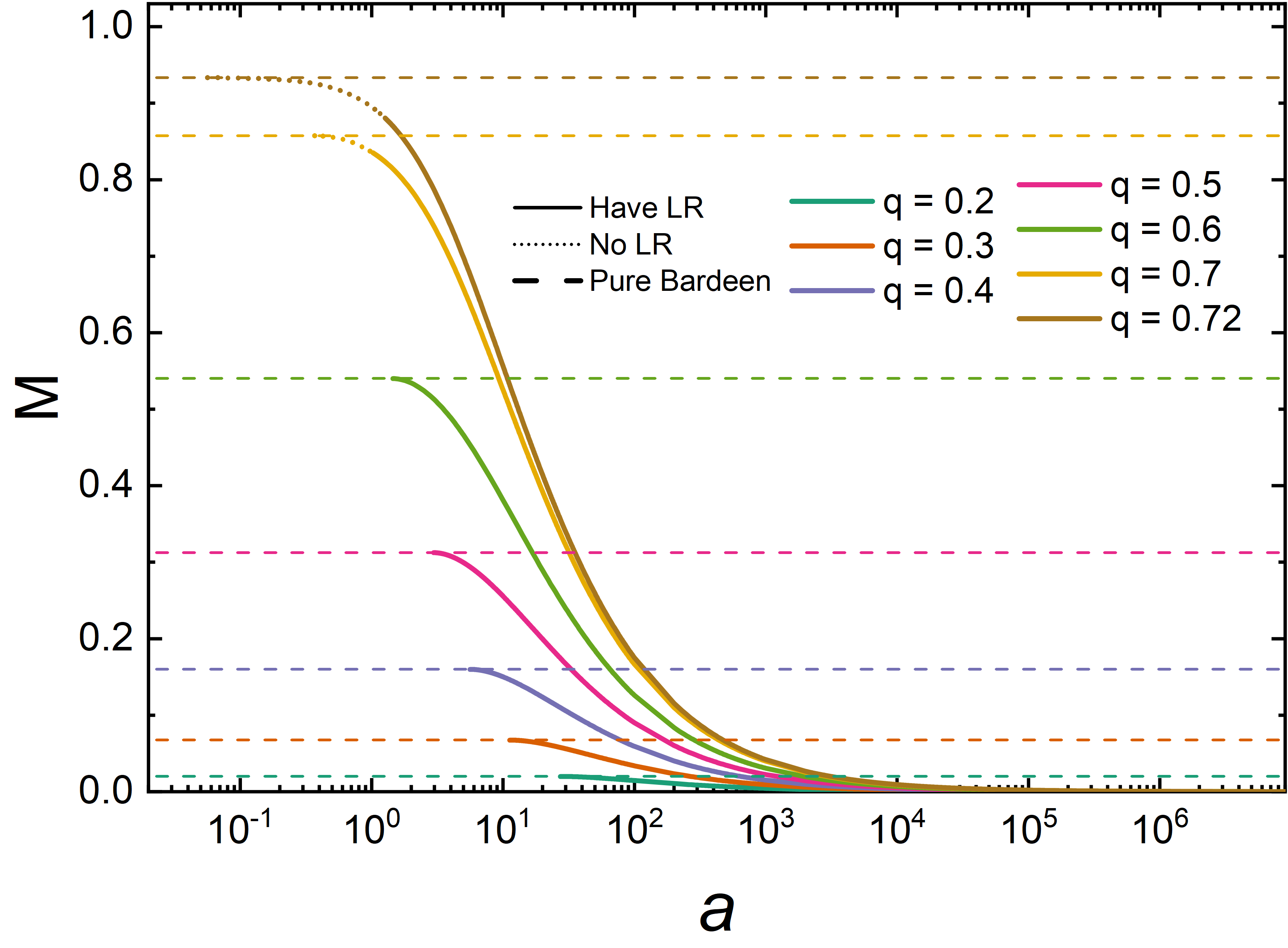}
			\label{fig:Mlittle}
		}	 
  		\subfigure{  
			\includegraphics[height=.28\textheight,width=.32\textheight, angle =0]{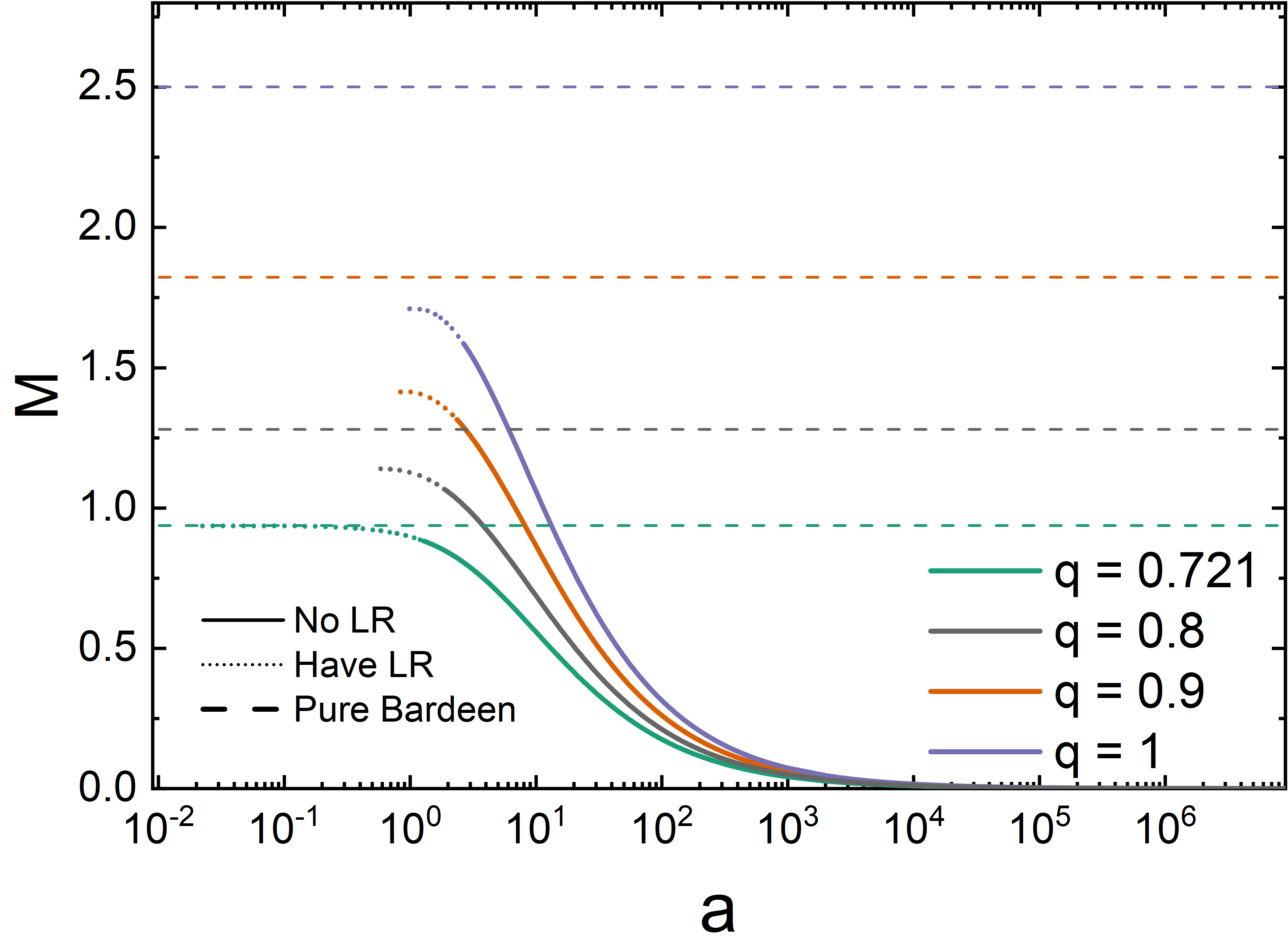}
			\label{fig:Mbig}
		}	
          		\subfigure{  
			\includegraphics[height=.28\textheight,width=.32\textheight, angle =0]{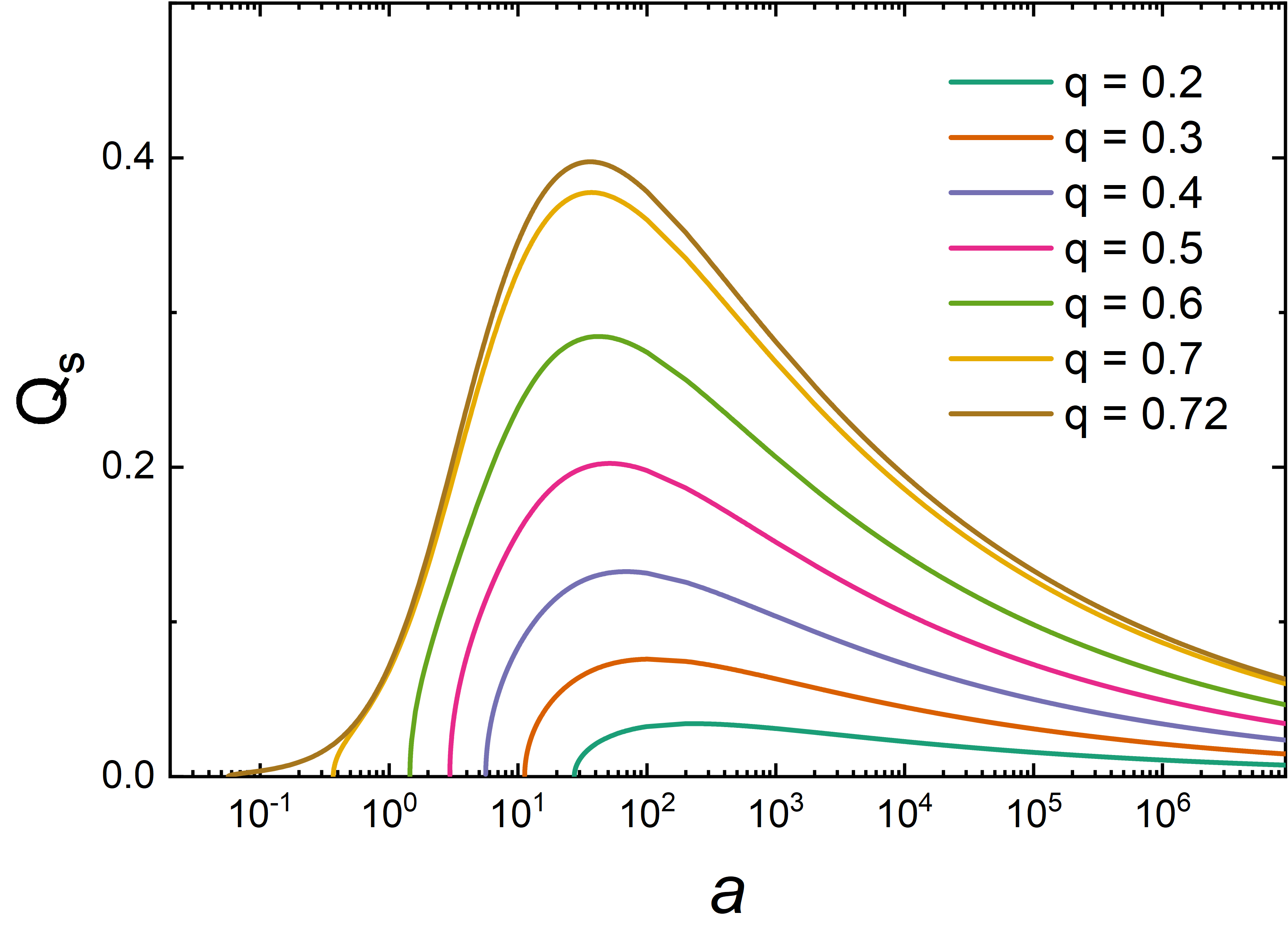}
			\label{fig:wlittle}
		}
          		\subfigure{  
			\includegraphics[height=.28\textheight,width=.32\textheight, angle =0]{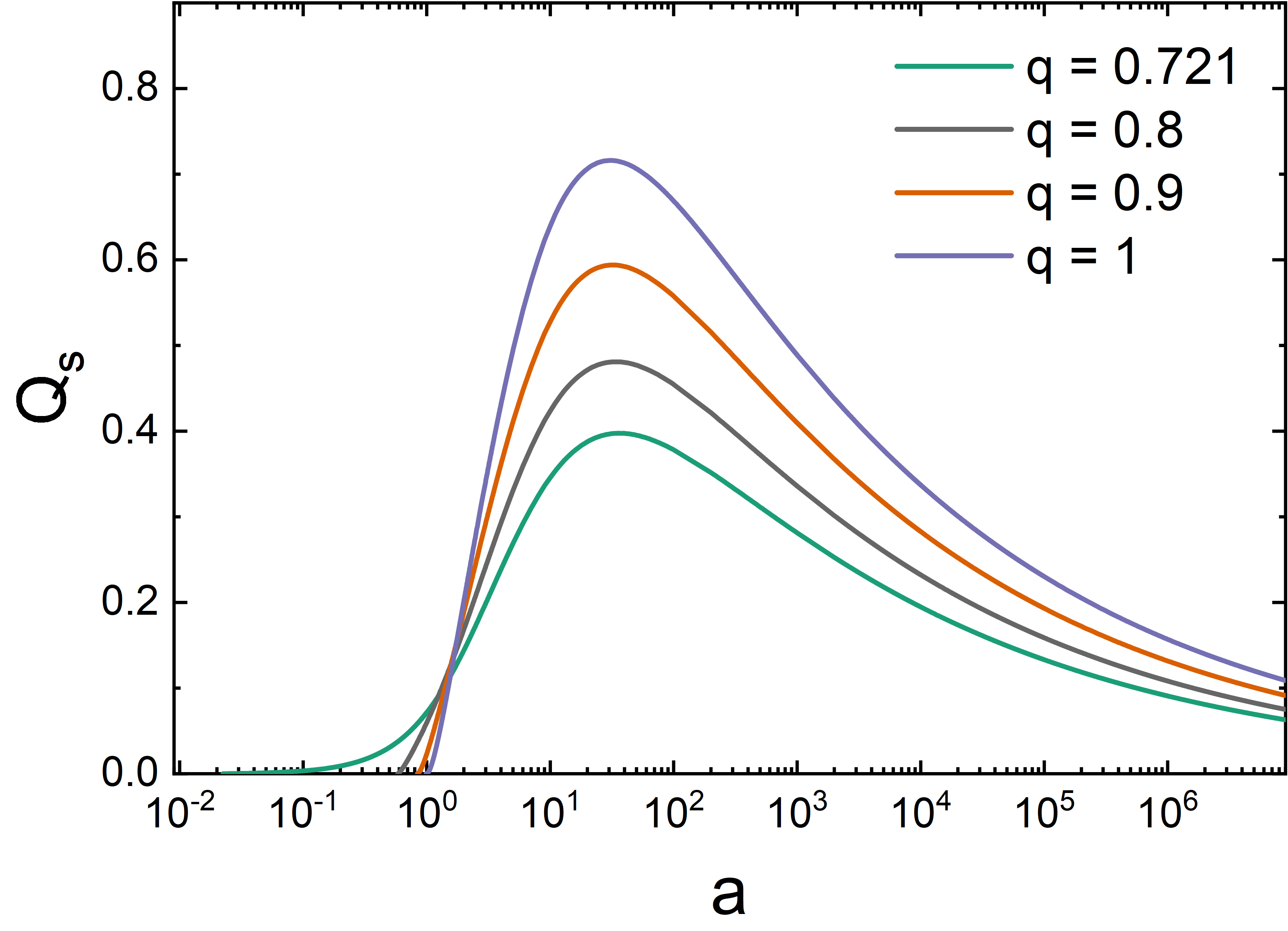}
			\label{fig:wbig}
		}        
  		\end{center}		
		\caption{ The mass $M$ (top) and scalar charge $Q_s$ (bottom) versus the coupling parameter $a$ with the different values of magnetic charges $q$. In the top panel, the dotted line indicates that the solution admits the light ring (LR), and the dashed line represents the pure Bardeen spacetime under the same parameter.}
	\label{fig:matterandcharge}	
		\end{figure}

The domain of existence of SBS can correspond to the curves in the scalar charge $Q_s$ (bottom panel) or ADM mass $M$ (top panel), versus coupling parameter $a$, diagram - Fig.~\ref{fig:matterandcharge}. The parameter $a$ corresponding to the left endpoints of these curves is the scalarization threshold $a_{t}$. 

For a fixed $q$, from Fig.~\ref{fig:matterandcharge}, it can be observed that there exists a maximum value for $Q_s$ and $M$, which increases as $q$ increases. However, the solutions corresponding to the maximum value of $M$ and $Q_s$ are not the same. For $Q_s$ since its value increases initially and then decreases, the maximum value of $Q_s$ is attained at a relatively larger value of $a$. In contrast, $M$ decreases monotonically with $a$, so the maximum value corresponds to the solution with the smallest coupling parameter $a$. A comparison between the left and right panels reveals that for magnetic charges smaller than $q_c=0.7208$, the maximum mass $M_{max}$ of the SBS solution is nearly the same as that of the pure Bardeen spacetime (dashed line) under same magnetic charges - see the left panel). 

	\begin{figure}[!htbp]
		\begin{center}
		\subfigure{ 
			\includegraphics[height=.28\textheight,width=.32\textheight, angle =0]{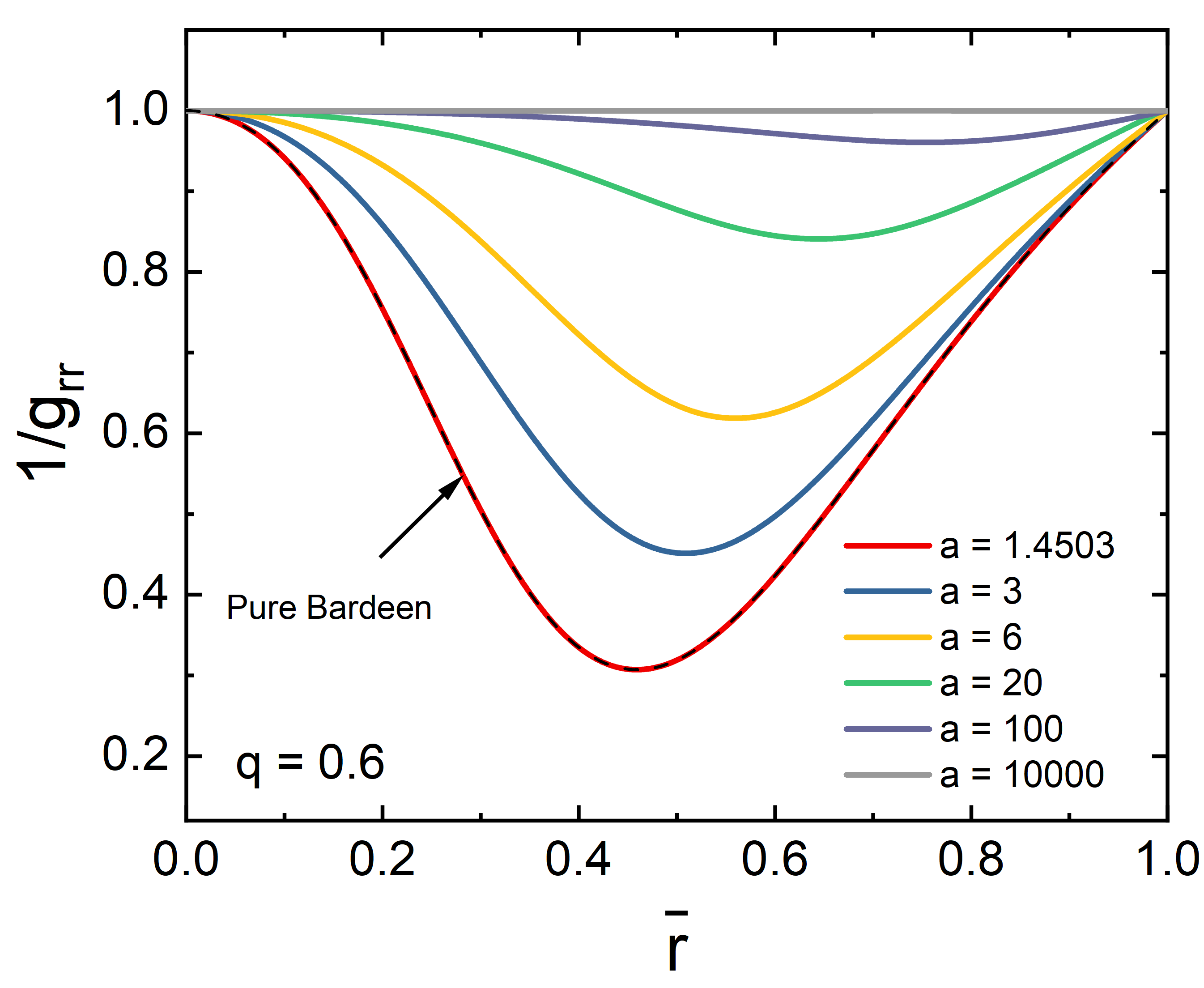}
			\label{fig:q06n}
		}	 
  		\subfigure{  
			\includegraphics[height=.28\textheight,width=.32\textheight, angle =0]{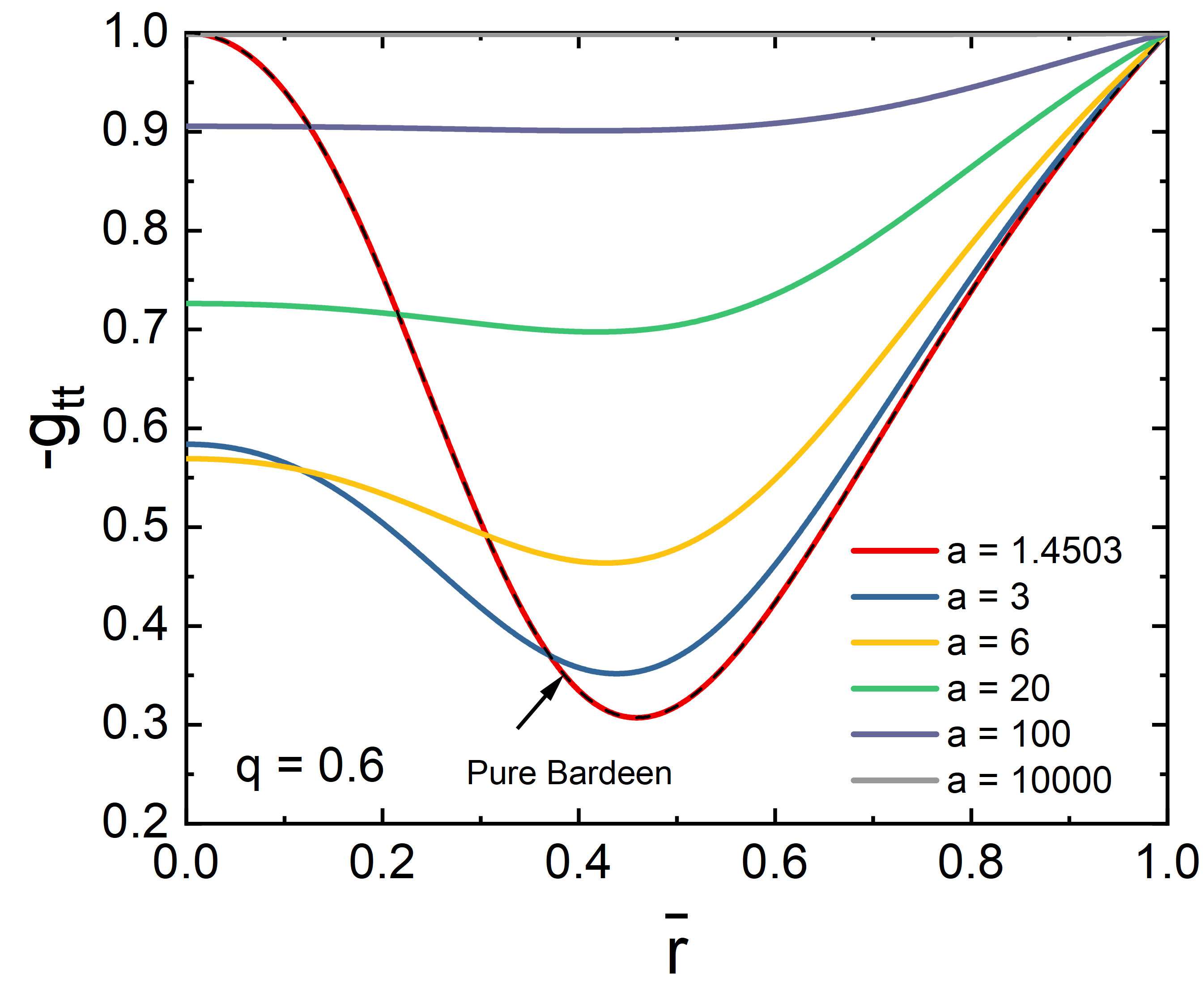}
			\label{fig:q06no2}
		}	
    \quad
        \subfigure{ 
			\includegraphics[height=.28\textheight,width=.32\textheight, angle =0]{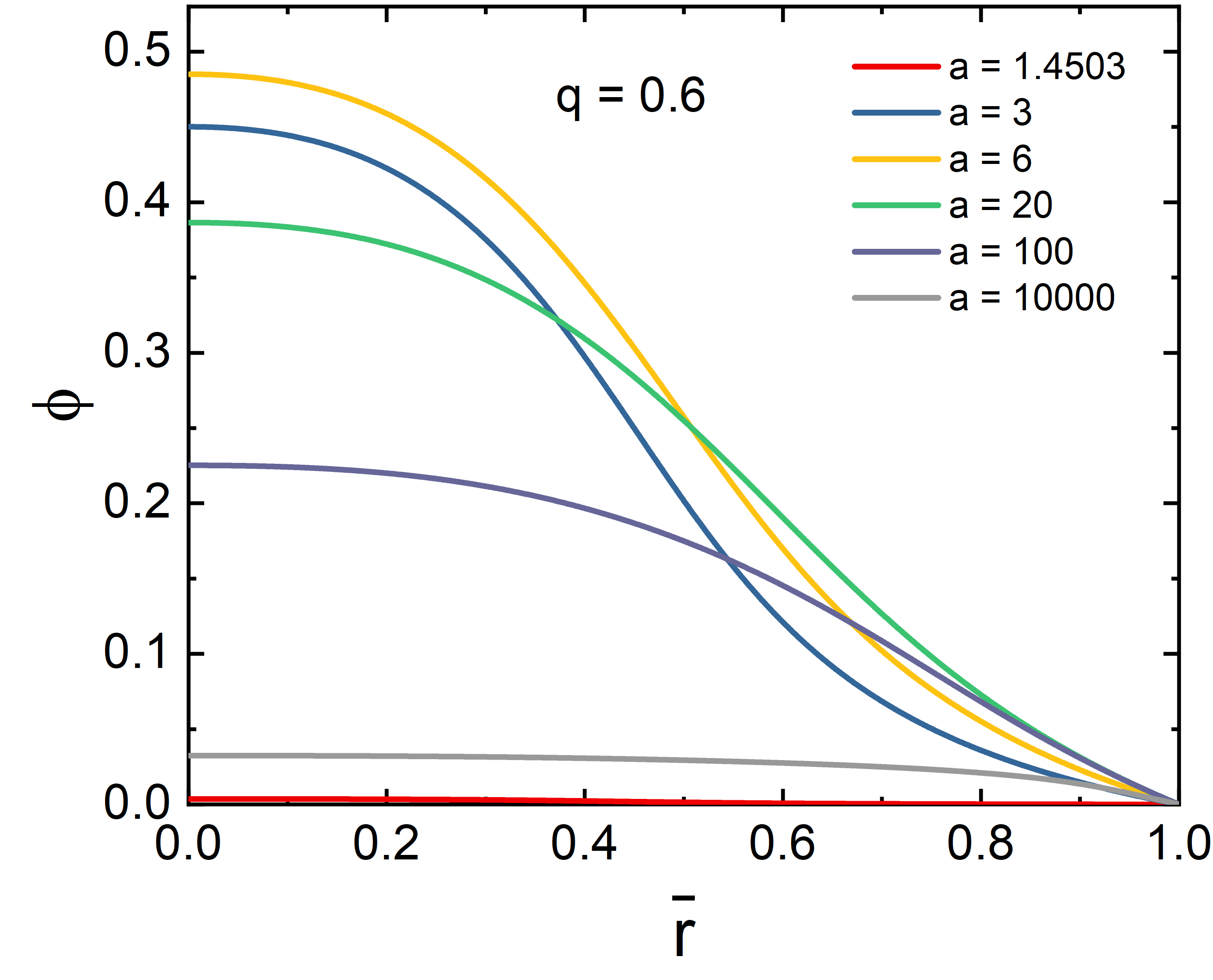}
			\label{fig:q06phi}
		}	 	
  		\end{center}	
		\caption{The metric field function $1/g_{rr}=n$, $-g_{tt}=n\sigma^2$ and real scalar field function $\phi$ as a function of $\bar{r}$ with $q=0.6$. The dashed line  represents the pure Bardeen spacetime.}
	\label{fig:q06function}	
		\end{figure}

However, once the magnetic charge exceeds $q_c$ (right panel), the mass of the Bardeen solution can significantly surpass $M_{max}$, and the difference between them becomes increasingly pronounced as the magnetic charge increases. This means that the introduction of the scalar field will decrease the ADM mass of the SBS and hence the mass of SBS is lower than that of pure Bardeen's. Moreover, the value $a_t$ decreases with $q$ for $q < q_c$, and increases for $q > q_c$. Therefore, in the following, we will discuss each of these two scenarios in turn.

	\begin{figure}[!htbp]
		\begin{center}
		\subfigure{ 
			\includegraphics[height=.28\textheight,width=.32\textheight, angle =0]{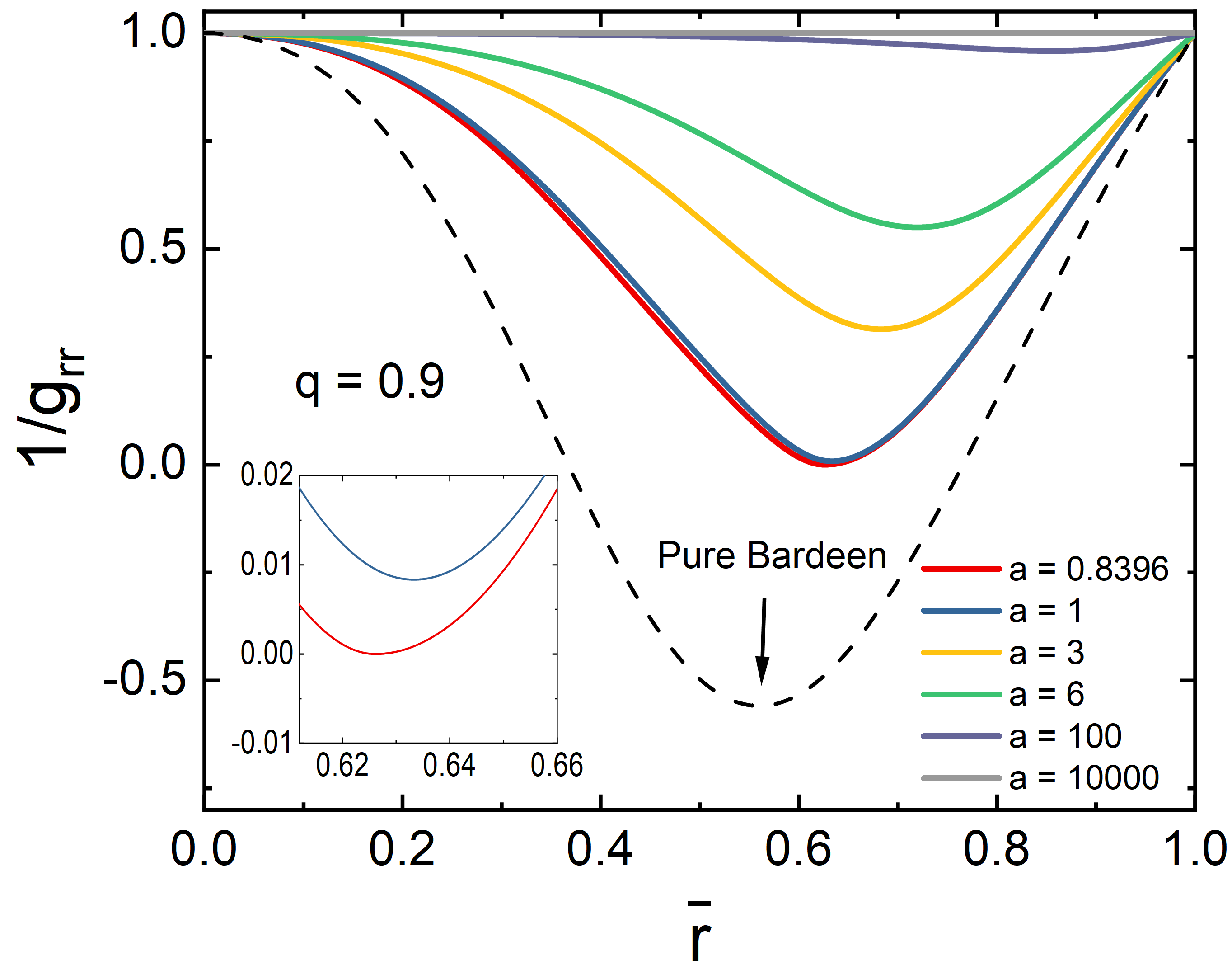}
			\label{fig:q09n}
		}	 
  		\subfigure{  
			\includegraphics[height=.28\textheight,width=.32\textheight, angle =0]{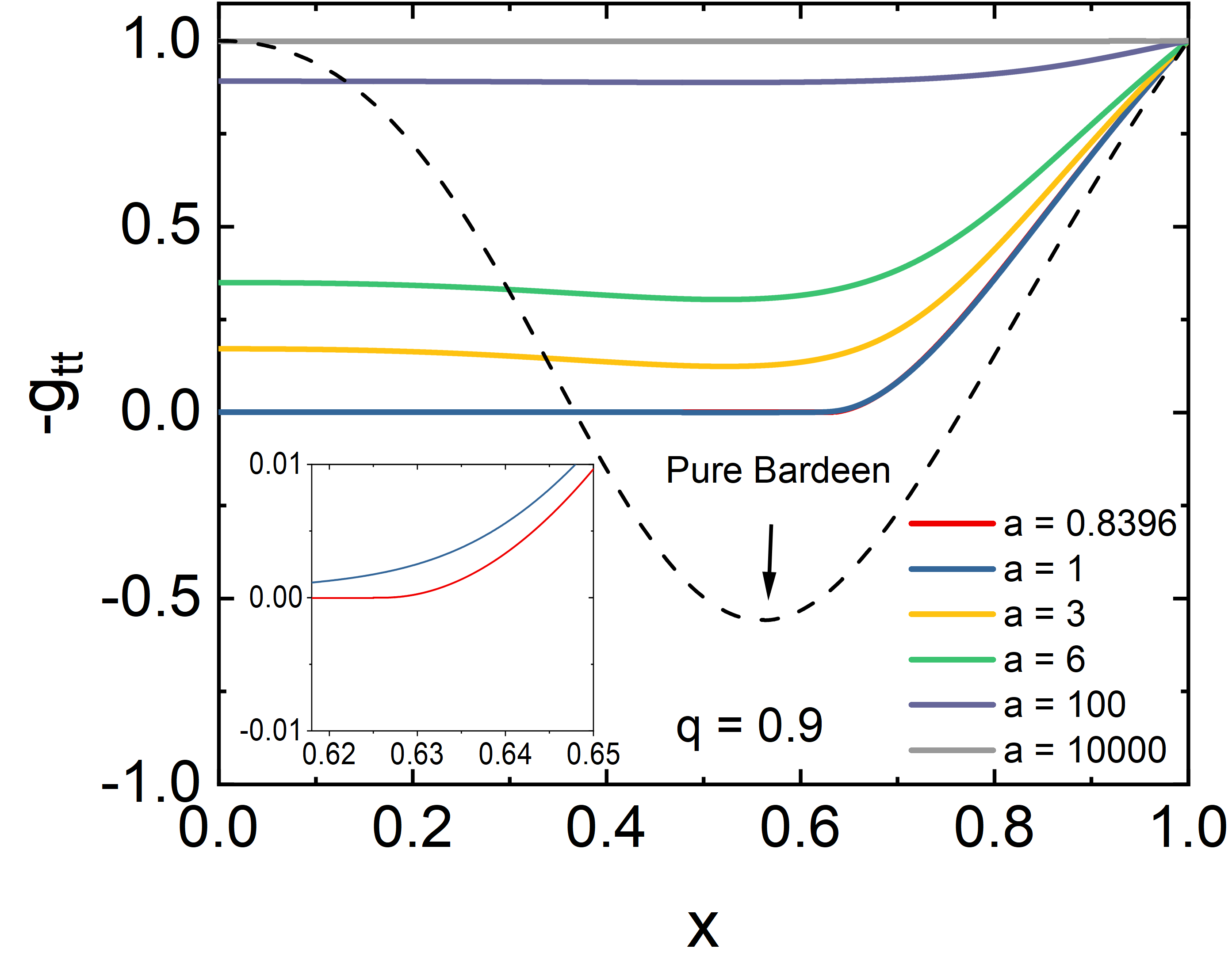}
			\label{fig:q09no2}
		}	
    \quad
        \subfigure{ 
			\includegraphics[height=.28\textheight,width=.32\textheight, angle =0]{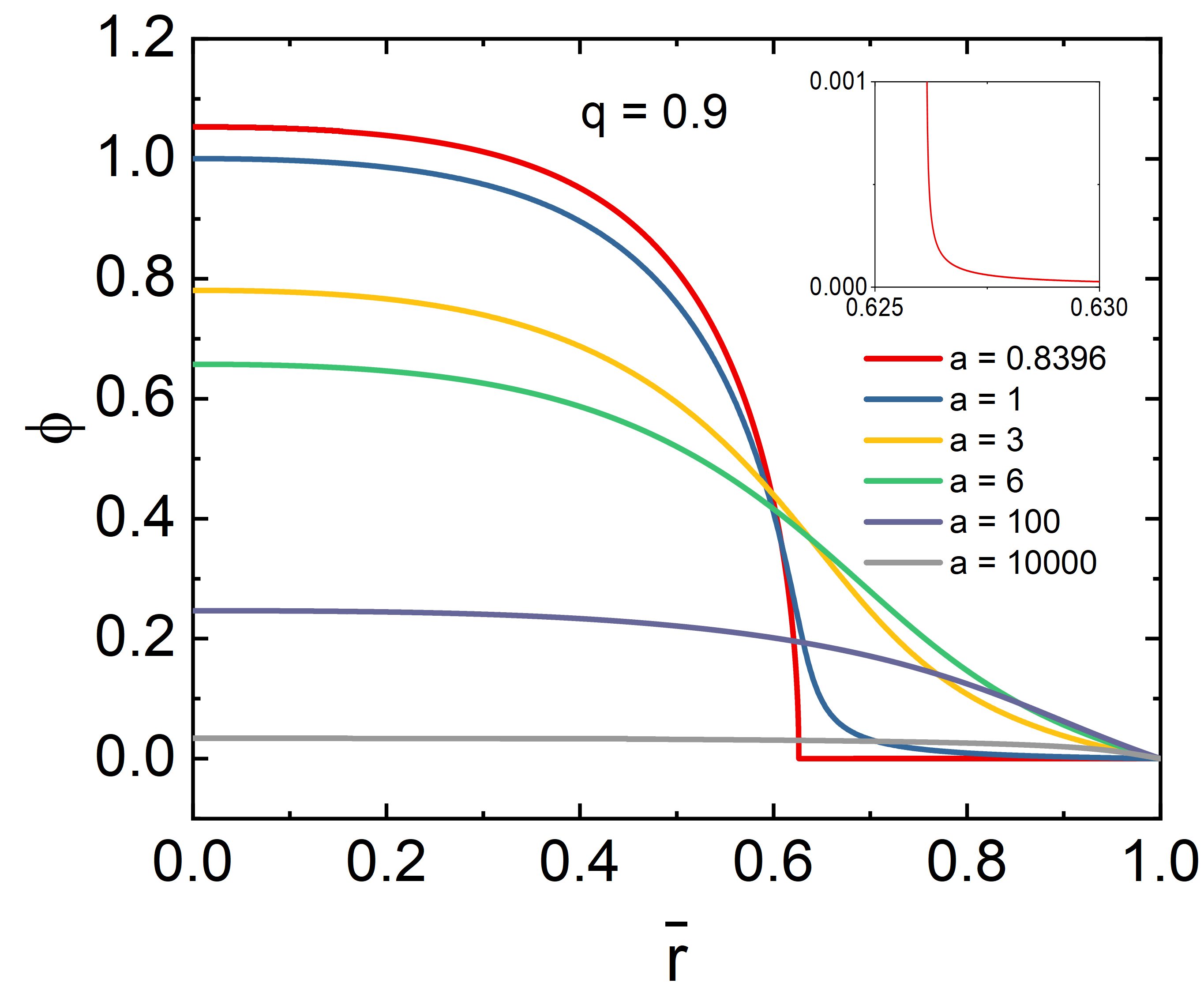}
			\label{fig:q09phi}
		}	 
  		\end{center}	
		\caption{The radial distribution of the field functions with $q = 0.9$.}
	\label{fig:q09function}	
		\end{figure}

\subsection{\texorpdfstring{$q < q_{c}$}{q < q_c}~ solution families}
In this subsection, we discuss the SBS in the case $q < q_c$. As a example, Fig.~\ref{fig:q06function} show the profiles of the metric field functions $-g_{tt}=no^2$, $1/g_{rr}=g^{rr}=n$, and the scalar field $\phi$ with several values of $a$ for $q=0.6$. It can be observed that in the case of $a=1.4503$ (nearly the scalarization threshold, see~Fig.~\ref{fig:matterandcharge}), the metric function closely resembles that of a pure Bardeen spacetime (black dashed line), with the scalar field nearly vanishing and can be regarded as a small perturbation. As the coupling parameter $a$ increases, the maximum of the scalar field function $\phi_{max}$ first grows and then decreases, while the minimum value (denote as $-g_{tt(min)}$ and $g^{rr}_{(min)}$) of the metric function $-g_{tt}$ and $g^{rr}$ gradually rises. In particular, when $a=10000$, the metric functions approach that of a nearly flat solution. Therefore, we conjecture that in the limit of $a \to \infty$, the spacetime of the SBS model becomes a flat spacetime.

\subsection{\texorpdfstring{$q \geq q_c$}{q \leq q_c}~ solution families}

In the case of $q>q_c$, the distributions of the field functions $-g_{tt}$, $1/g_{rr}$, and $\phi$ with several values of magnetic charge are shown in Fig.~\ref{fig:q09function}. In contrast to the other case, the maximum value of the scalar field function varies monotonically with $a$. As $a$ approaches $a_t$, the scalar field will not become very diluted, cannot be regarded as a small perturbation, and a special solution can emerge. For this solution, as seen from the red line in Fig.~\ref{fig:q09function}, the scalar field becomes almost entirely confined within a certain position $r_{cH}$. 

	\begin{figure}[!htbp]
		\begin{center}
		\subfigure{ 
			\includegraphics[height=.28\textheight,width=.33\textheight, angle =0]{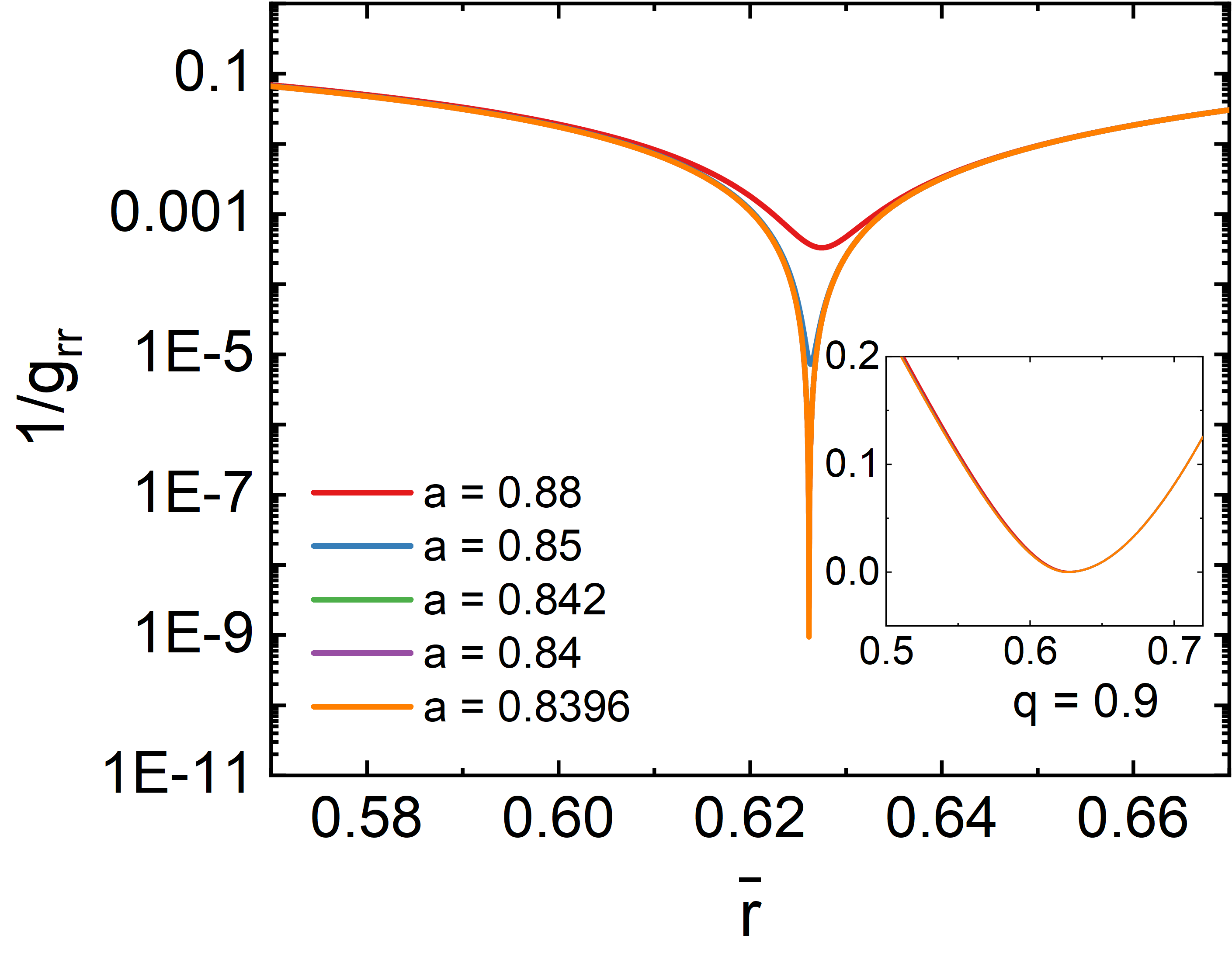}
		}	 
  		\subfigure{  
			\includegraphics[height=.28\textheight,width=.33\textheight, angle =0]{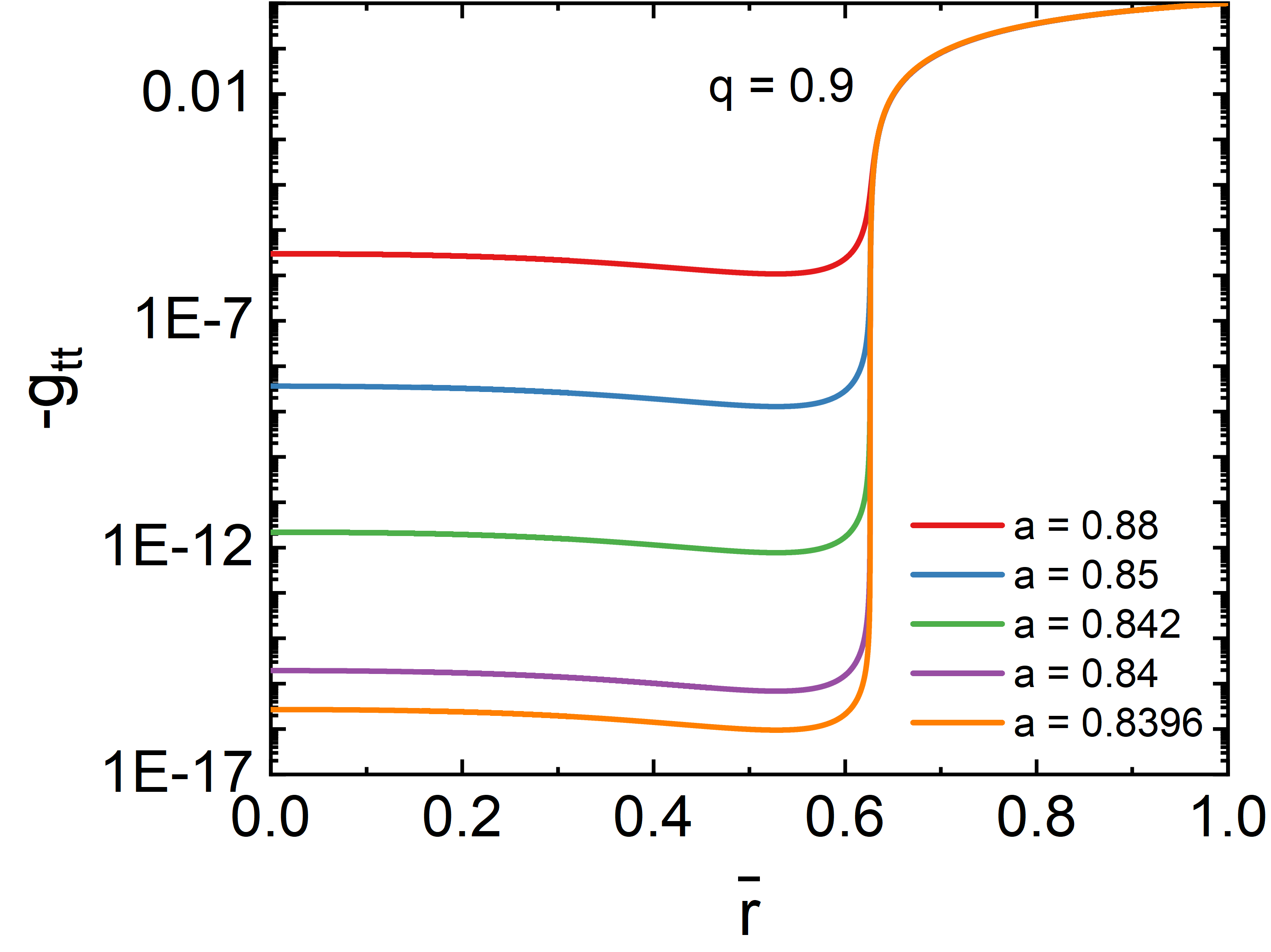}
		}	        
  		\end{center}	
		\caption{The metric function $1/g_{rr}$ and $-g_{tt}$ with $a\rightarrow a_t$ and $q = 0.8$.}
	\label{fig:logfunction}	
		\end{figure}

Moreover, as the parameter $a$ approaches $a_{t}$, the metric function $-g_{tt}$ attains a minimum value that becomes very close to zero (the order of $\mathcal{O}(10^{-16})$) within $r_{cH}$, while the metric function $g^{rr}$ tends to almost vanish at $r_{cH}$ (the order of $\mathcal{O}(10^{-9})$) - see Fig.~\ref{fig:logfunction}. That is, objects inside the critical horizon move exceedingly slowly for an observer at infinity, as if ``frozen”. It is noteworthy that because $g^{rr}$ is not exactly zero, the solution does not possess an event horizon. Hence, it is not classified as a black hole but rather a ``frozen” SBS. Given that the character of $r_{cH}$ is a very close analogy with the horizon of BHs, the position $r_{cH}$ can be termed as ``critical horizon". 

In fact, as long as $a > a_{t}$, the ``frozen" phenomenon becomes more evident as $a$ gets closer to $a_{t}$\footnote{That is, within the accuracy of our computations, when $a \le a_t$, the scalar field vanishes and the solution is a pure Bardeen spacetime; 
and when $a > a_t$, there exists a positive parameter $e = a - a_t$, such that as $e \to 0$, the minima of $-g_{tt}$ and $1/g_{rr}$ can be made arbitrarily close to zero from the perspective of numerical computation (but never exactly zero).
}. However, considering numerical computations and for convenience in discussion, without loss of generality, we only call the solution corresponding to the first value of $a$ beyond $a_{t}$ (i.e., the left end point of the curves in Fig.~\ref{fig:matterandcharge}), where the frozen phenomenon is most evident, as the frozen SBS, with $a$ rounded to four decimal places.

\subsection{Orbit of the photon}

Photon orbits can produce some observable effect in astronomy, and studying them in a given spacetime is crucial for understanding the properties of that spacetime. Thus, we will discuss the orbit of the test photon in SBSs, which is ruled by the following null geodesics equation 
\begin{equation}
    g_{\mu\nu}\dot{x}^{\mu}\dot{x}^{\nu}=0
    \label{eq:geod}
\end{equation}
Here, dots denote derivatives with respect to the affine parameter $\lambda$ along the geodesic. Since the static,  spherically symmetric spacetime, without loss of generality, we assume the orbit of the photon lies on the equatorial plane $\theta=\pi/2$. In addition, the spacetime described by the metric (\ref{eq:metric}) possesses two Killing vector $\partial_t$ and $\partial_{\varphi}$, which correspond to two conserved quantities: the energy $E = -g_{tt}\dot{t}$ and angular momentum $L=g_{\varphi \varphi}\dot{\varphi}$. Therefore, Eq.~(\ref{eq:geod}) becomes
\begin{equation}
    \dot{r}^2+V(r)=0,
    \label{eq:rdgeod}
\end{equation}
where
\begin{equation}
    V(r)=\frac{1}{g_{tt}g_{rr}}\left(E^2+g_{tt}\frac{L^2}{r^2}\right). \label{eq:V}
\end{equation}
To determine the photon trajectories in the spacetime (\ref{eq:metric}), we need to numerically solve Eq.~(\ref{eq:rdgeod}). Following \cite{Luminet:1979nyg}, introducing the substitution $u = 1/r$, then Eq.~(\ref{eq:rdgeod}) can be transformed to:
\begin{equation}
    (\frac{du}{d\varphi})^2+\frac{1}{g_{tt}g_{rr}}\left(g_{tt}u^2+\frac{1}{b^2}\right)=0
    \label{eq:twouphi}
\end{equation}
with the impact parameter $b=L/E$. Differentiating (\ref{eq:twouphi}) with respect to $\varphi$ and substituting metric (\ref{eq:metric}) yields
\begin{equation}
    \frac{d^2u}{d\varphi^2}=\frac{o'}{b^2u^2o^2}+\frac{n'}{2}-un
\end{equation}
Then, numerical integration of this equation can determine the orbit of the photon.

\begin{figure}[t]
\begin{center}
\subfigure[$V_{\text{eff}}$]  {\label{fig:eff07}
\includegraphics[width=3.9cm,height=3.9cm]{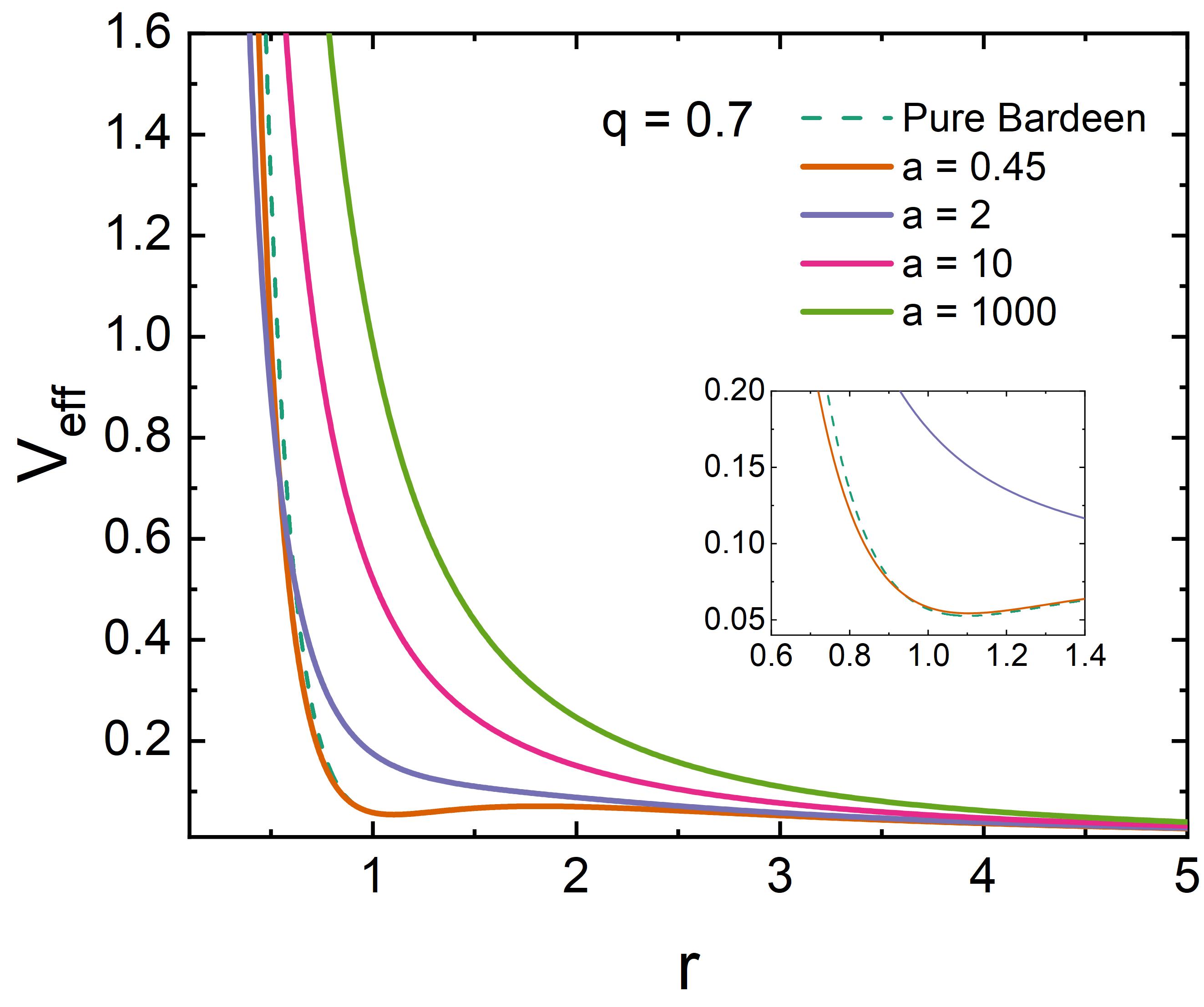}}
\subfigure[$a=0.45$]  {\label{fig:UOq07a045}
\includegraphics[width=3.8cm,height=3.8cm]{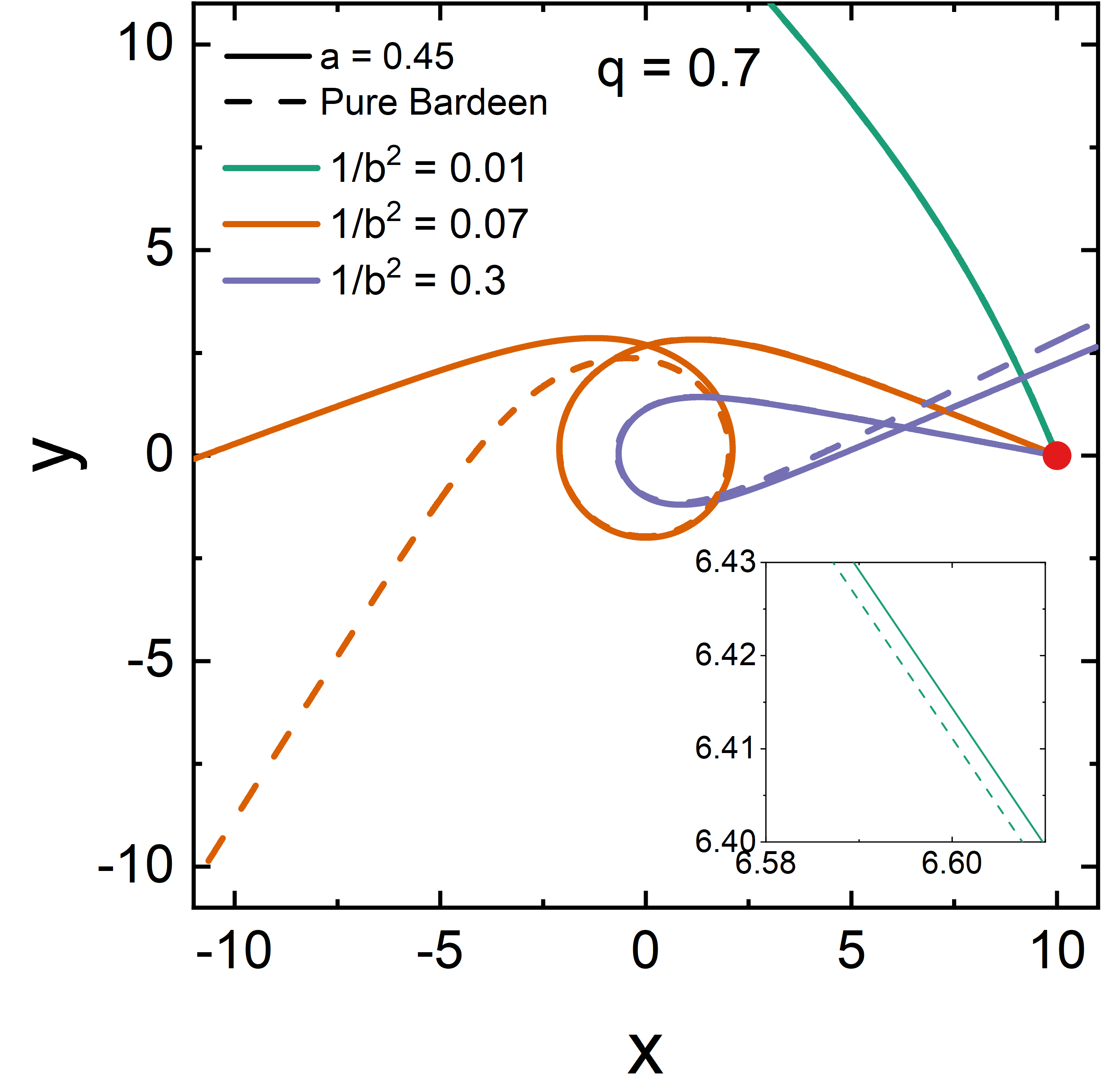}}
\subfigure[$a=2$]  {\label{fig:UOq07a2}
\includegraphics[width=3.8cm,height=3.8cm]{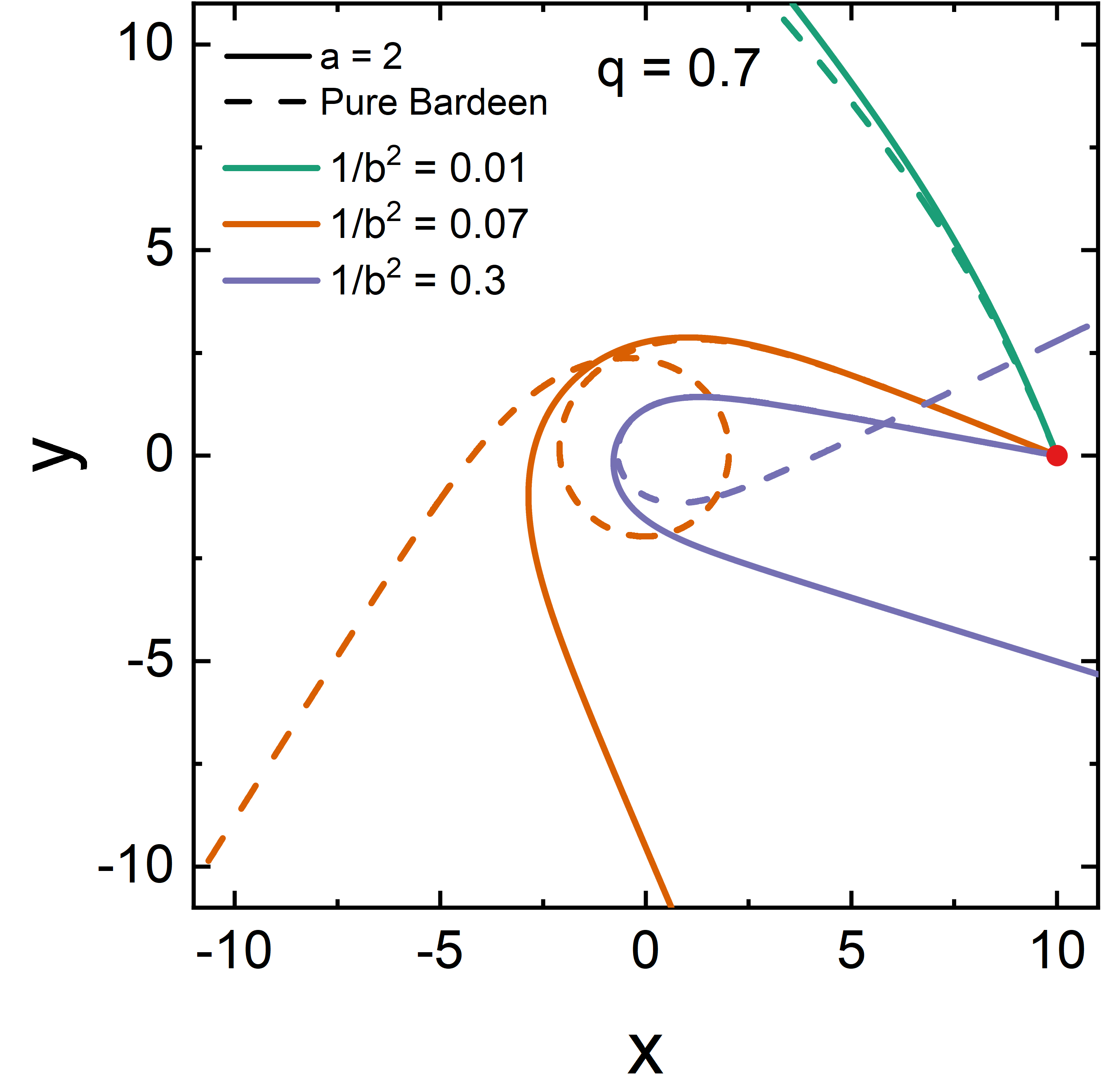}}
\subfigure[$a=10$]  {\label{fig:UOq07a10}
\includegraphics[width=3.8cm,height=3.8cm]{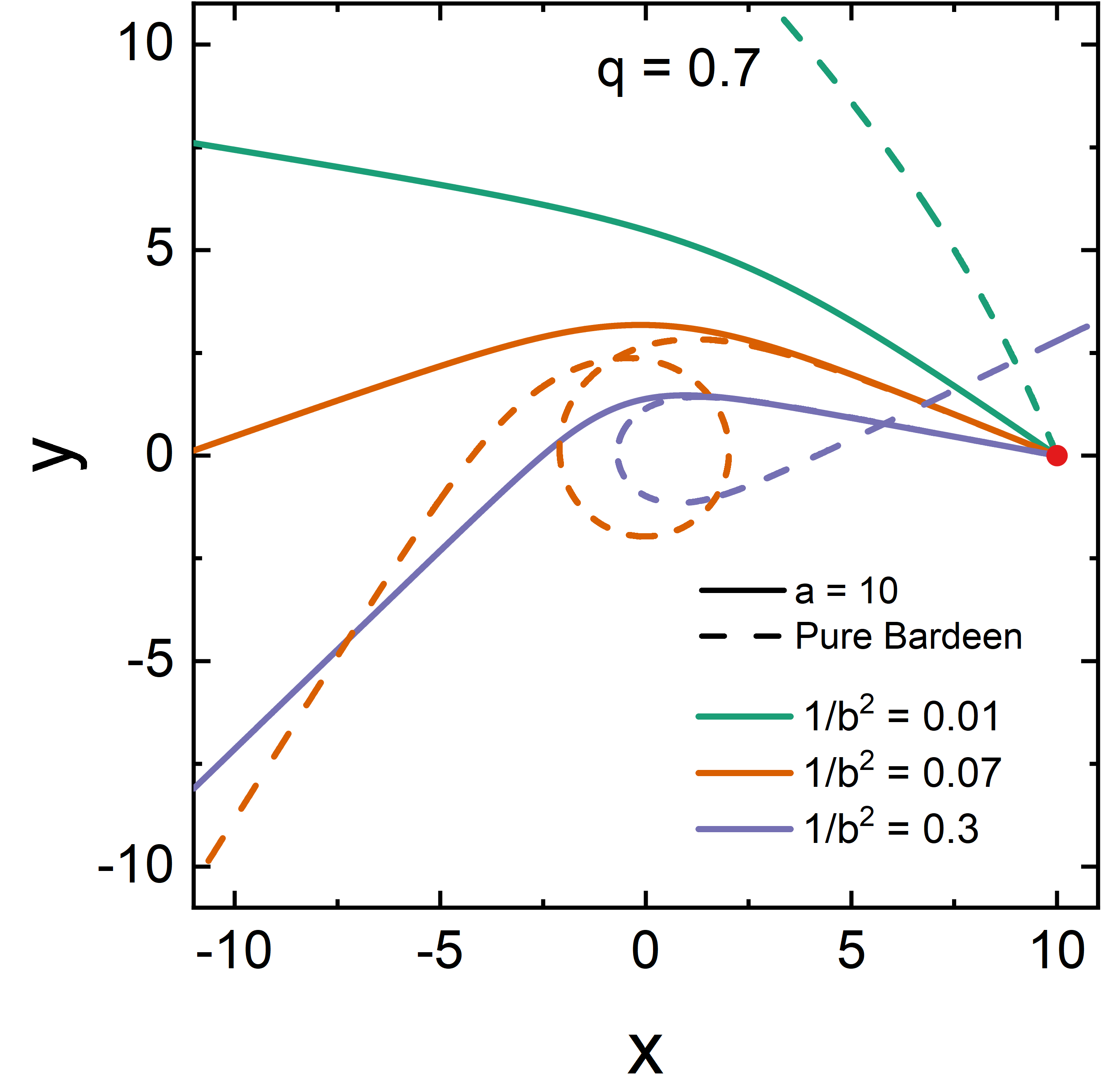}}\\
\end{center}
\caption{(\ref{fig:eff07}): The shapes of the effective potential $V_{\text{eff}}$ for different $a$ with $q=0.7$. (\ref{fig:UOq07a045}-\ref{fig:UOq07a10}): The corresponding unbound orbits in the background of SBSs (solid line) and pure Bardeen spacetime (dashed line) with the same parameter.}
\label{fig:unboundg}
\end{figure}

It is worth mentioning that the effective potential method is an effective way to study the motion of a classical test particle in a central field. From Eq.~(\ref{eq:rdgeod}) and Eq.~(\ref{eq:V}), the radial effective potential of a photon can be defined as:
\begin{equation}
    V_{\text{eff}}(r)=\frac{g_{tt}}{r^2}=\frac{no^2}{r^2},
\end{equation}
From Eq.~(\ref{eq:rdgeod}), we can deduce that, when $V'_{\text{eff}}=0$ and $V_{\text{eff}}=1/b^2$, the corresponding orbit is a circular orbit~\cite{Macedo:2013jja} (i.e., a light ring). Specifically, if $V_{\text{eff}}''<0$, the light ring (LR) is unstable, whereas if $V_{\text{eff}}''>0$, the light ring is stable~\cite{Cunha:2017qtt}.

As an example, the radial effective potentials for different parameters with $q=0.7$ ($q<q_c$) are illustrated in Fig.~\ref{fig:eff07}.
As shown in this figure, for $a=0.45$, the effective potential possesses two extreme points, indicating that the corresponding solution under this parameter configuration features two light rings: the inner one is stable while the outer one is unstable. Fig.~\ref{fig:matterandcharge} shows the domain of existence of SBSs with LRs for several magnetic charges. It can be seen that LRs appear when the scalarization parameter $a$ is relatively small.

In addition, from Fig.~\ref{fig:eff07}, it can also be observed that when $a$ is small, the effective potential of the SBS almost coincides with that of the Bardeen solution for the same parameters $s$ and $q$. As $a$ increases, the difference between them becomes more pronounced. Therefore, for the same particles in the background of SBSs and pure Bardeen solution with the same initial state, the difference in their orbits would become larger as $a$ increases, which can be seen clearly through Fig.~\ref{fig:UOq07a045},~\ref{fig:UOq07a2}, and~\ref{fig:UOq07a10} as well. 

\begin{figure*}[t]
\begin{center}
\subfigure[$V_{\text{eff}}$]  {\label{fig:eff08}
\includegraphics[width=4.2cm,height=3.8cm]{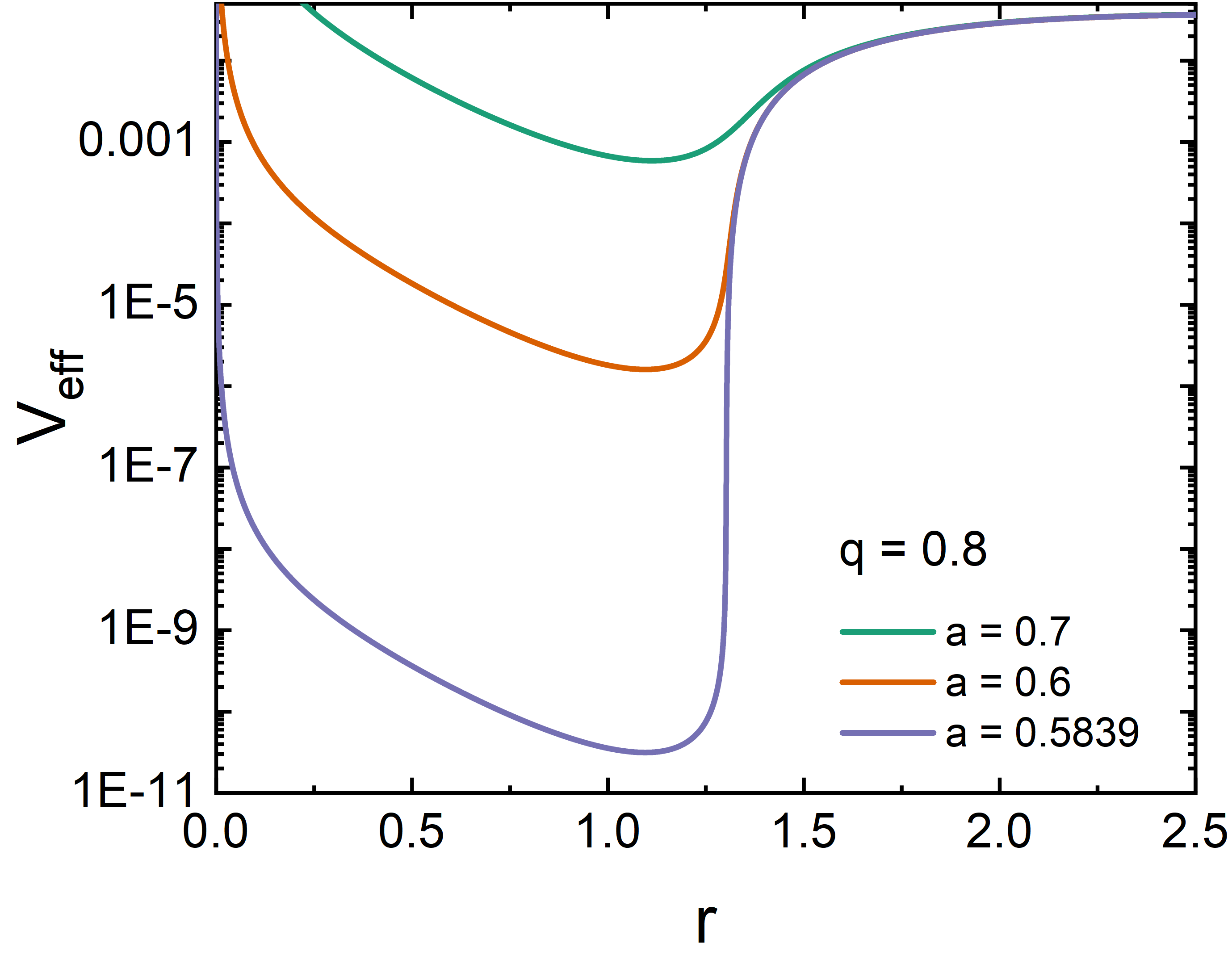}}
\subfigure[Frozen SBS]  {\label{fig:UBFOrbitq08a05839}
\includegraphics[width=3.8cm,height=3.8cm]{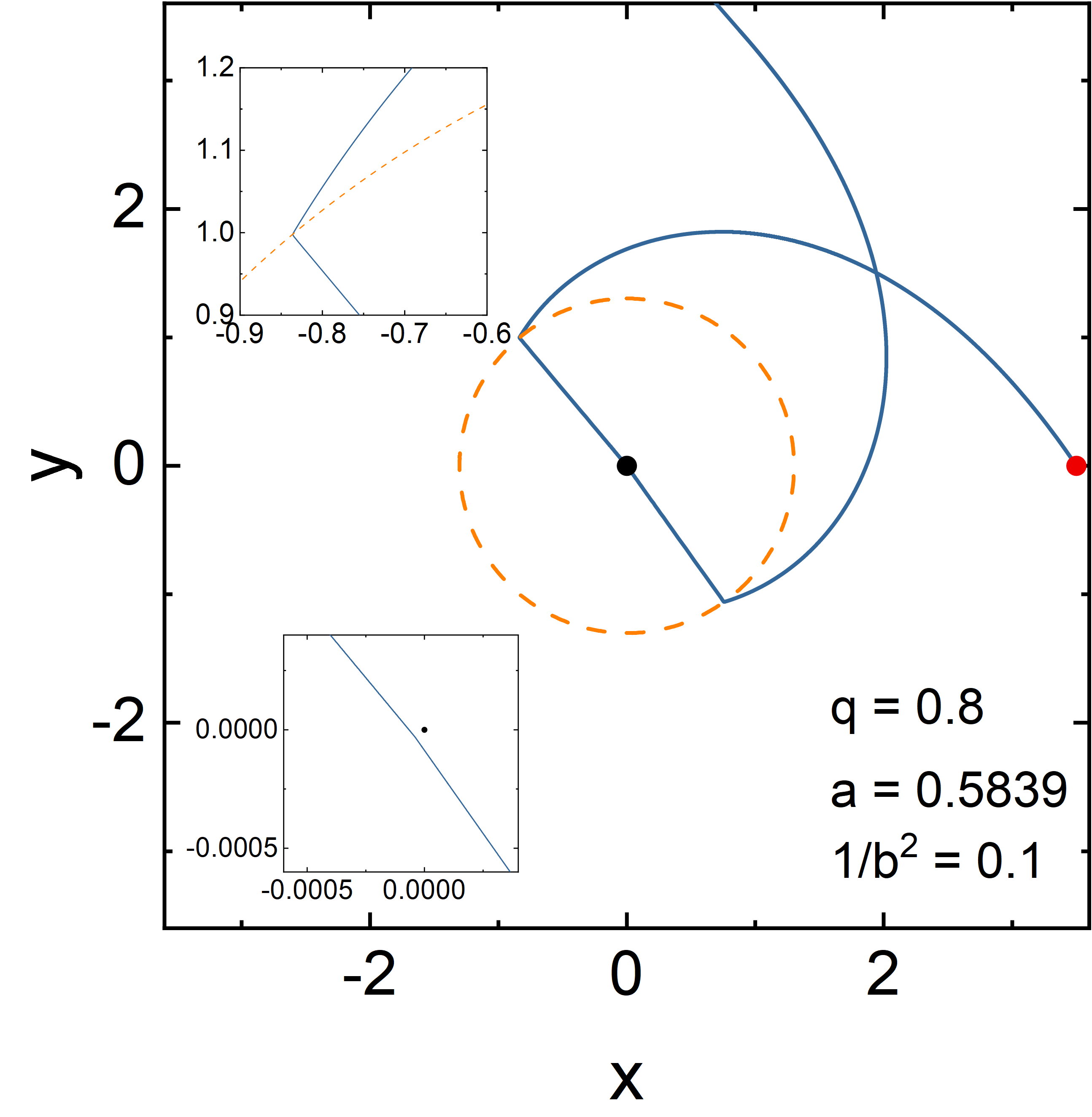}}
\subfigure[$a=0.6$]  {\label{fig:UBFOrbitq08a06}
\includegraphics[width=3.8cm,height=3.8cm]{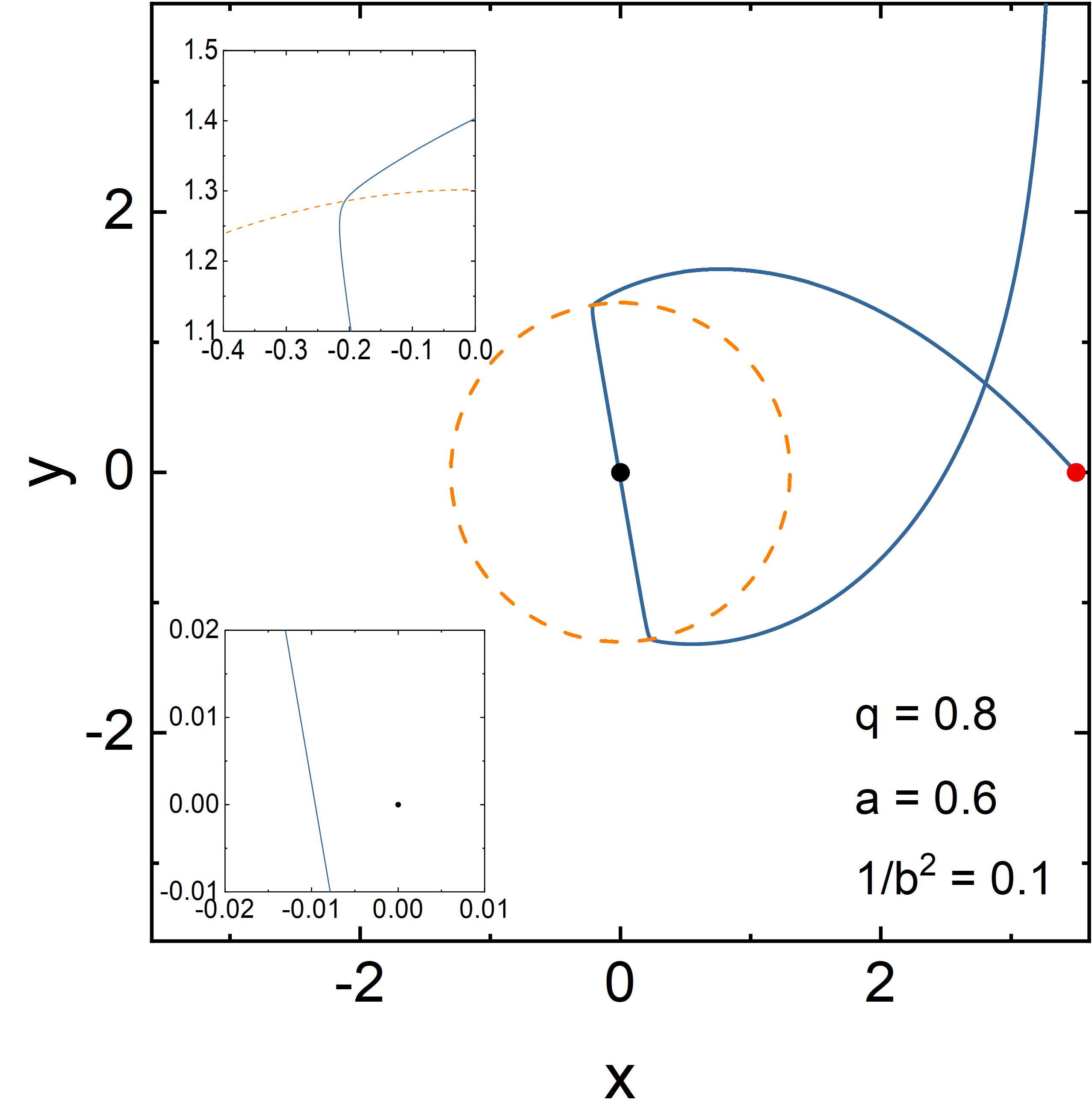}}
\subfigure[$a=0.7$]  {\label{fig:UBFOrbitq08a07}
\includegraphics[width=3.8cm,height=3.8cm]{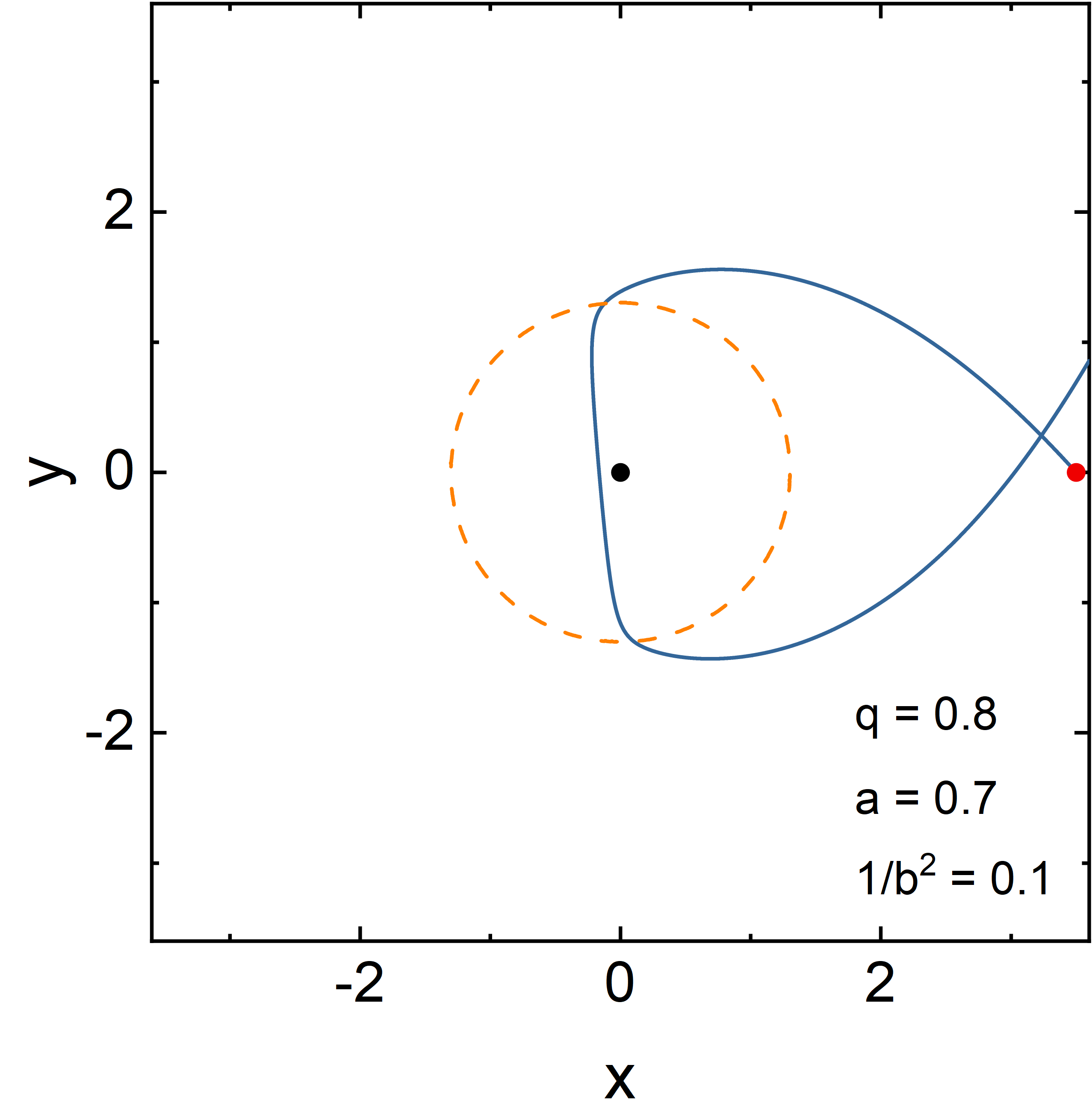}}\\
\end{center}
\caption{The effective potential $V_{\text{eff}}$ and unbound orbits around SBSs with several $a\rightarrow a_{t}$ for $q=0.8$. The dashed line represent the critical horizon.}
\label{fig:unboundf}
\end{figure*}

In the case of $q>q_{c}$, due to the distinctive nature, the orbital behavior of the frozen SBS is markedly different from that of other trajectories. To illustrate this, we show the orbit of SBS for several coupling parameters $a$ with $q=0.8$ in Fig.~\ref{fig:unboundf}, where $x=r\cos\theta$ and $y=r\sin\theta$. The dashed line represents the critical horizon. The corresponding radial effective potentials are presented in Fig.~\ref{fig:eff08}. It can be observed that when $a=0.7$, the deflection of the photon’s trajectory is relatively small. As $a$ decreases, the deflection increases. Eventually, when $a=0.5839$, the frozen state solution, the orbit undergoes a very large deflection near the critical horizon, and inside the critical horizon, it almost turns into a straight line and can pass extremely close to the origin (0,0). 
	\begin{figure}[!htbp]
		\centering	
        \subfigure[]{  
			\includegraphics[height=.20\textheight,width=.21\textheight, angle =0]{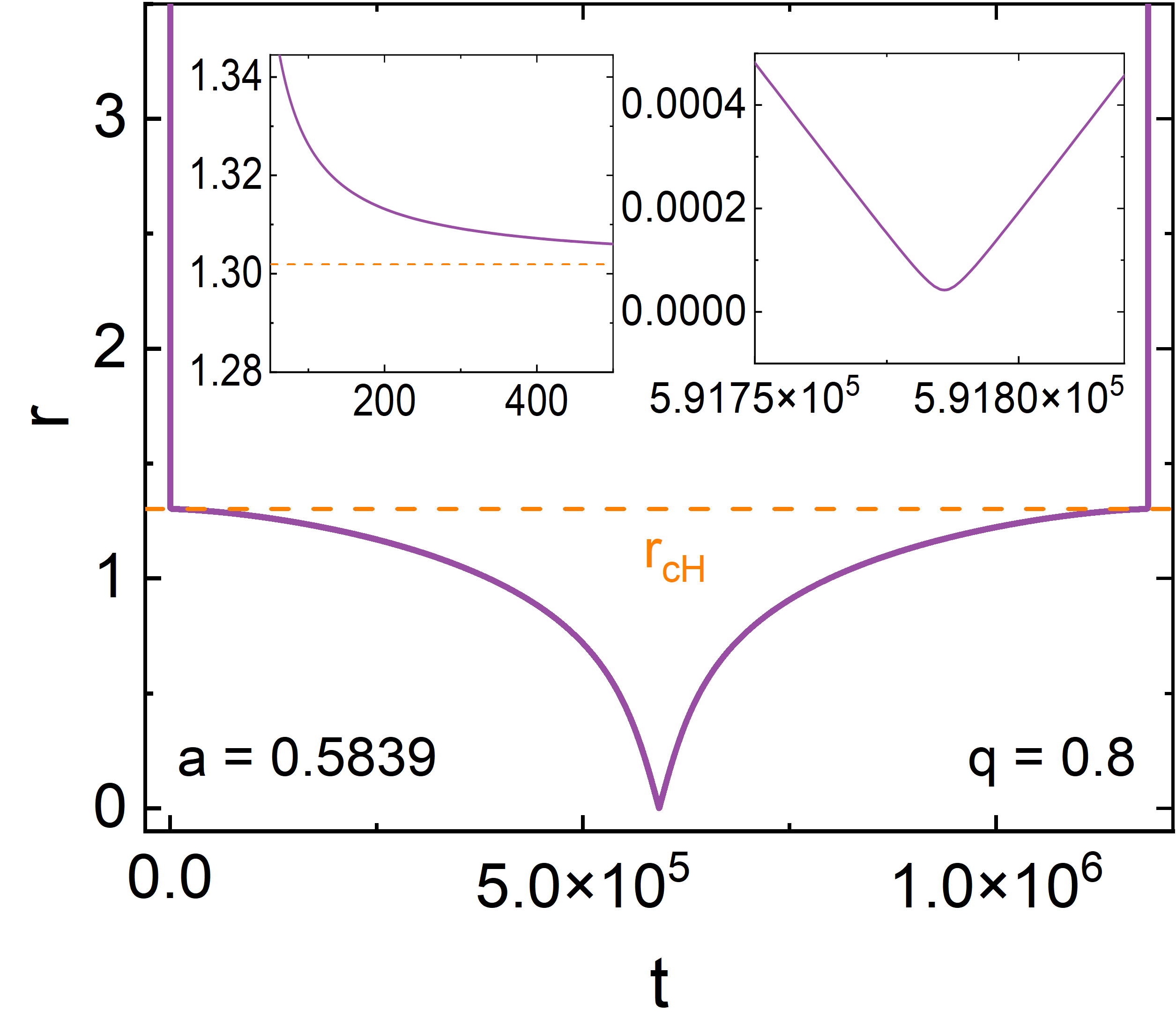}\label{fig:rt5839}
		}
		\subfigure[]{
			\includegraphics[height=.20\textheight,width=.21\textheight, angle =0]{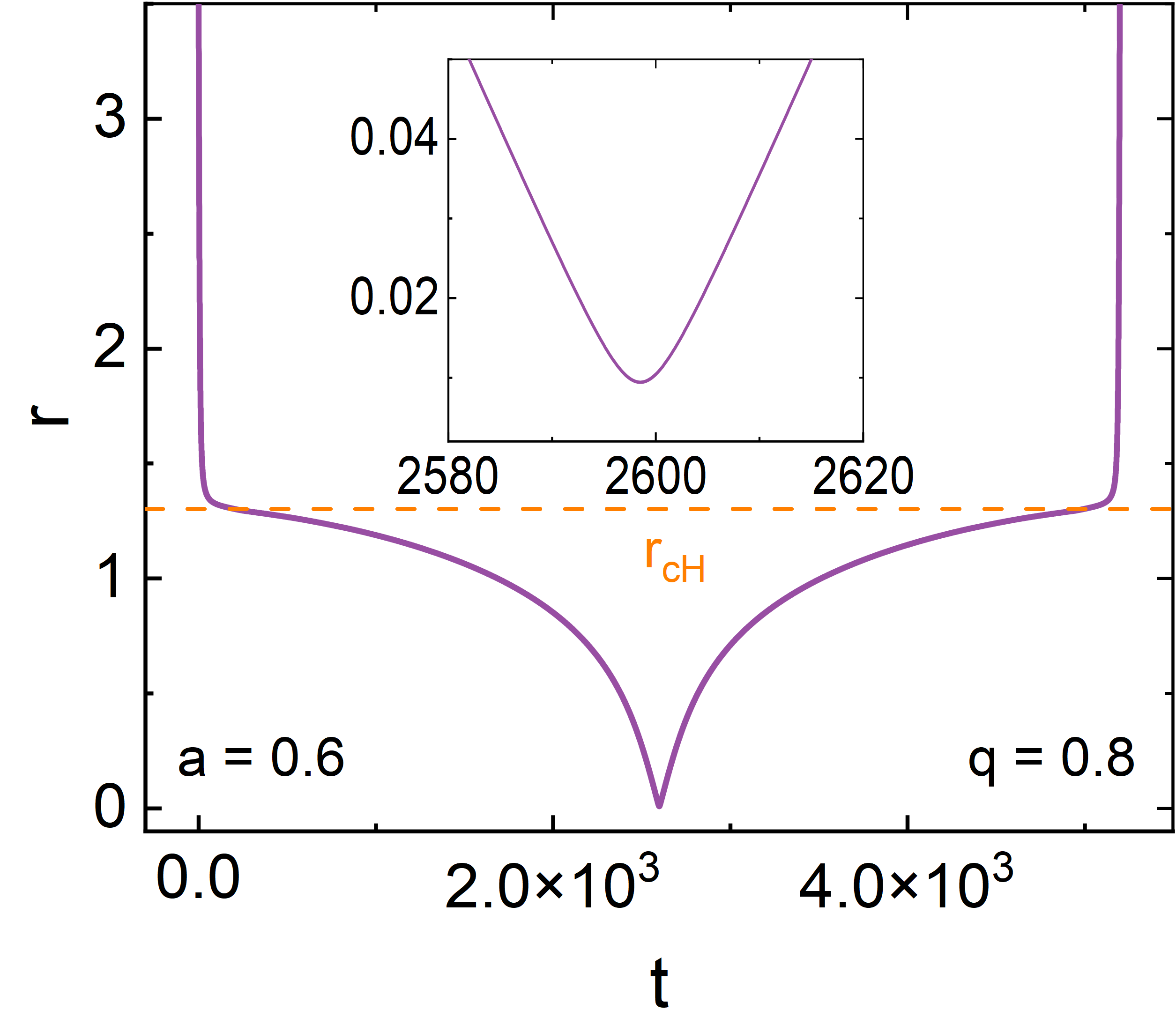}\label{fig:rt06}
   		} 	
            \subfigure[]{
			\includegraphics[height=.20\textheight,width=.21\textheight, angle =0]{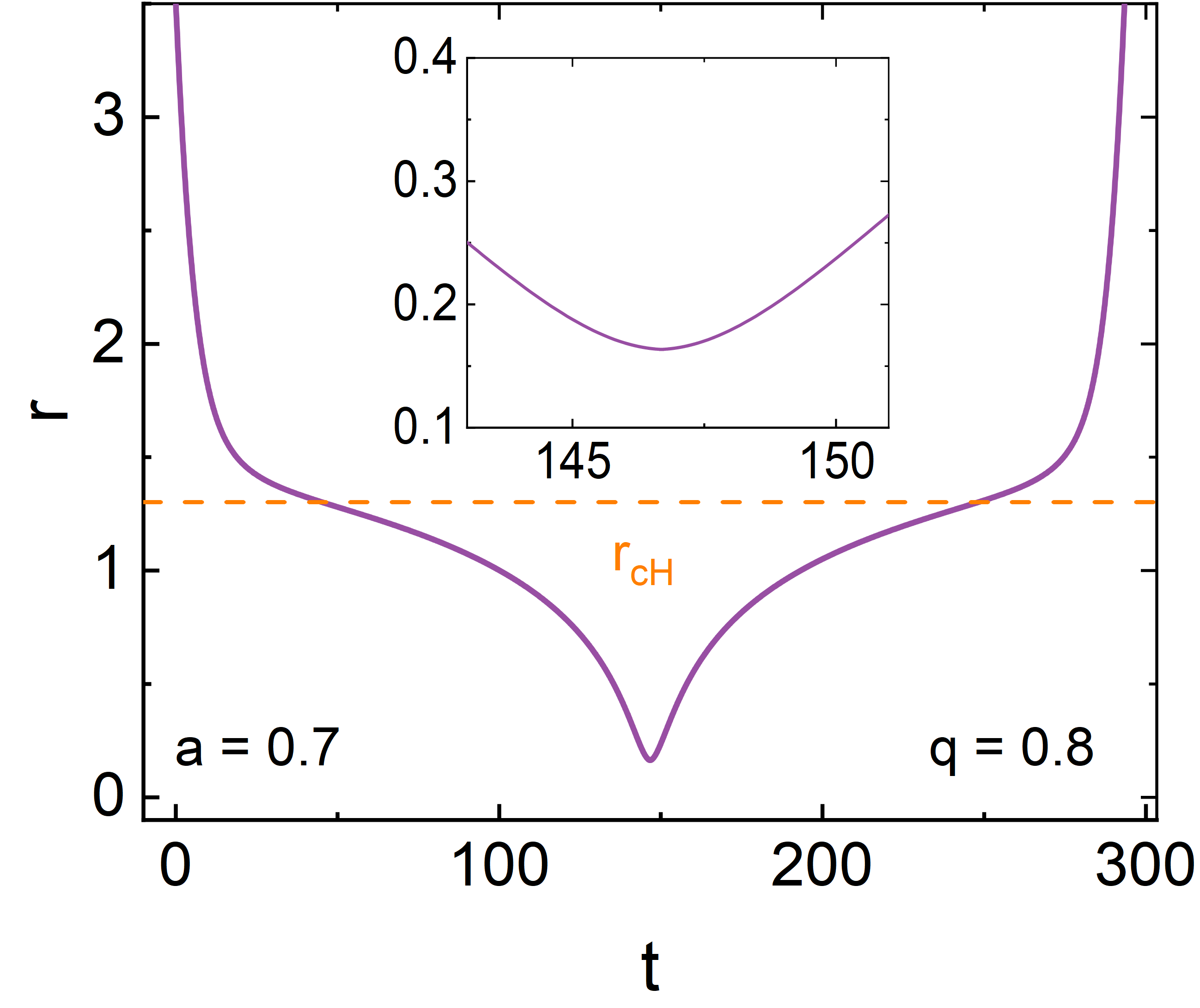}\label{fig:rt07}
   		} 
		\caption{The relationship between radial distance $r$ and time $t$ of the null geodesic orbit in the background of SBS.}
		\label{fig:rt}		
		\end{figure}

The evolution of the radial distance of test photons in the background of the frozen SBS is shown in Fig.~\ref{fig:rt}. The 
initial state of phton is $(t,r)=(0,3.5)$. By observing the results of this figure, we can notice that as $a$ approaches $a_{t}$, the test photons spend an exceptionally long time moving inside the critical horizon, particularly in its vicinity.

	\begin{figure}[!htbp]
\begin{center}
\subfigure{
\includegraphics[width=3.8cm,height=3.8cm]{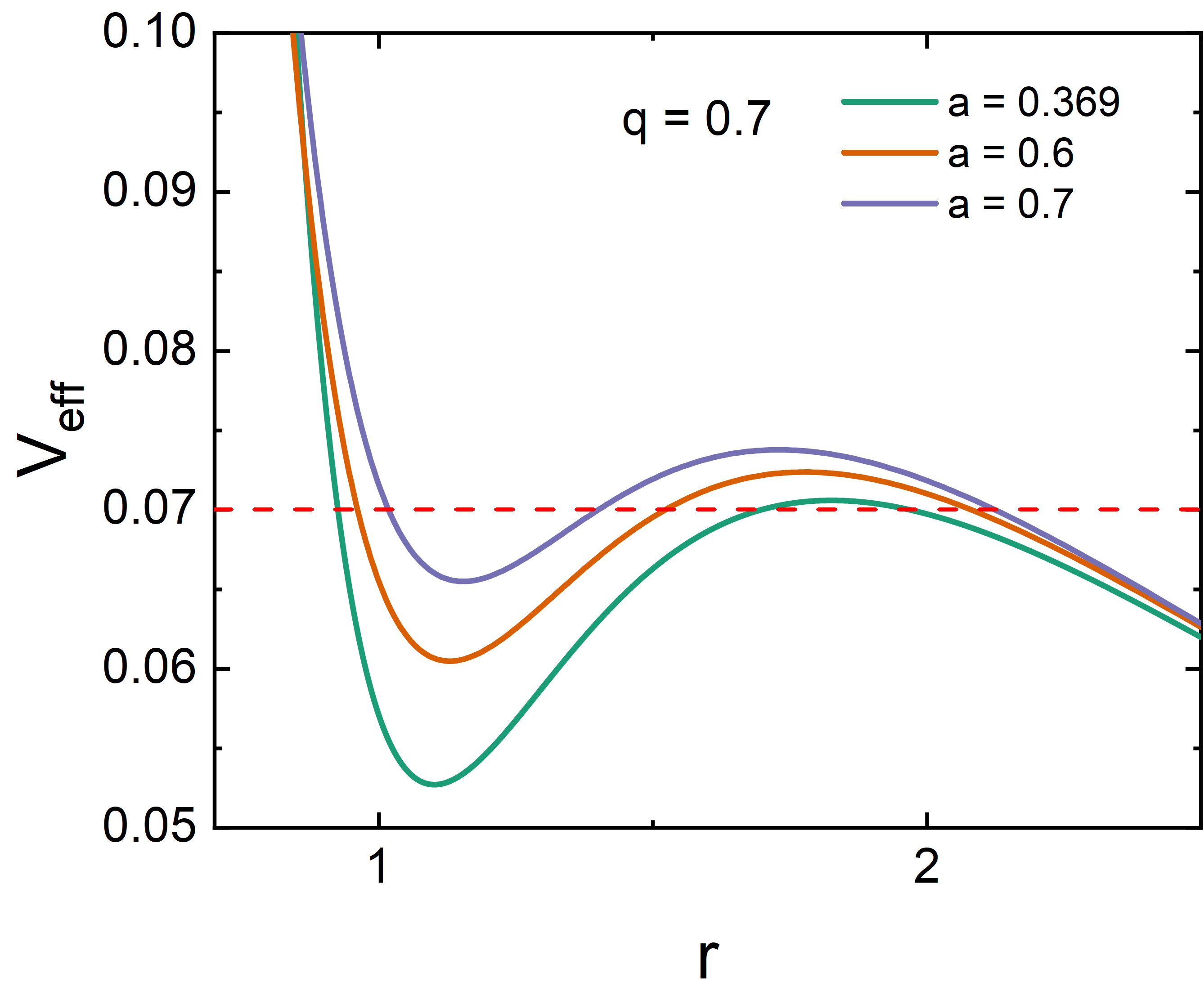}}
\subfigure{
\includegraphics[width=3.8cm,height=3.8cm]{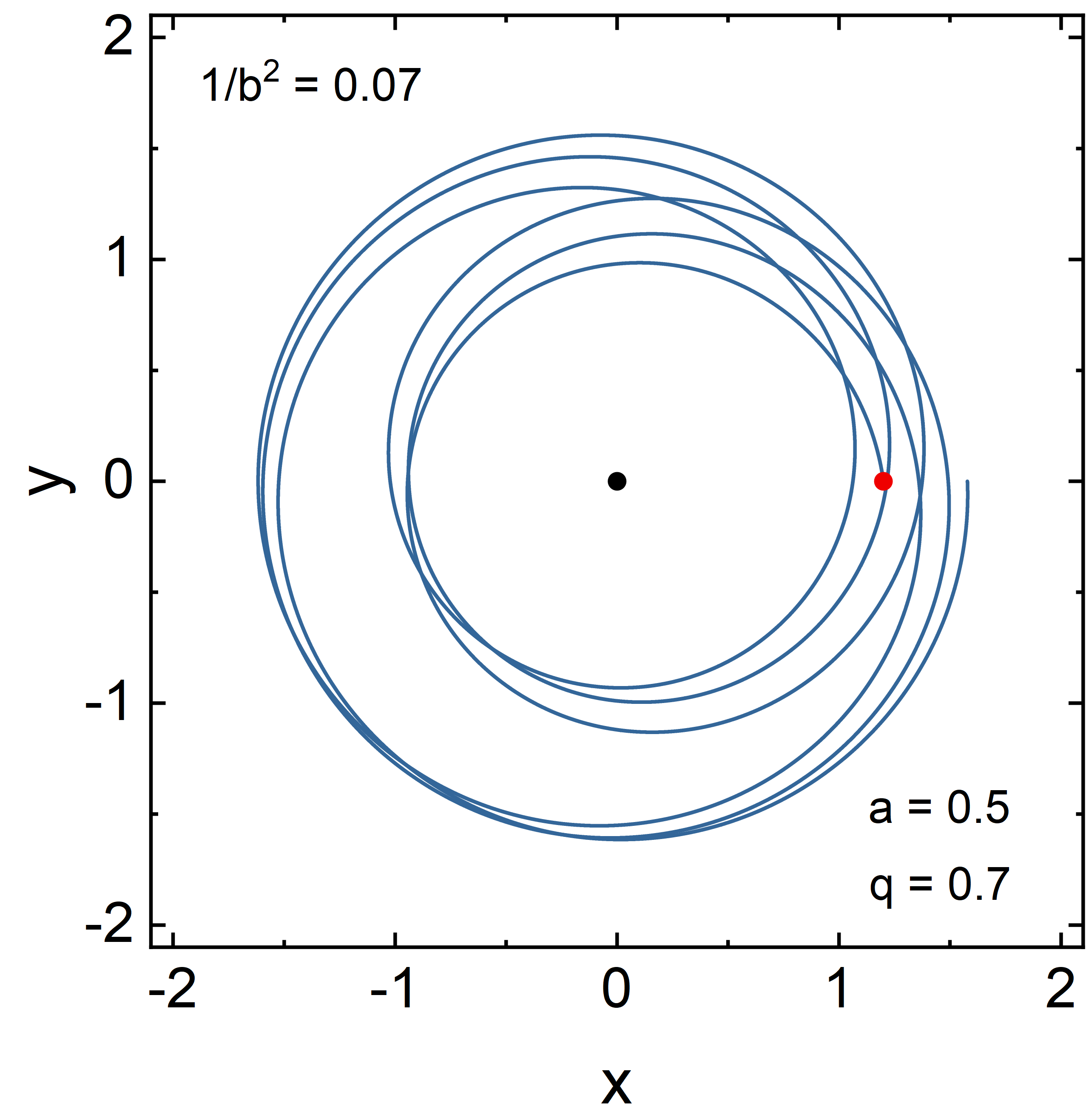}}
\subfigure{
\includegraphics[width=3.8cm,height=3.8cm]{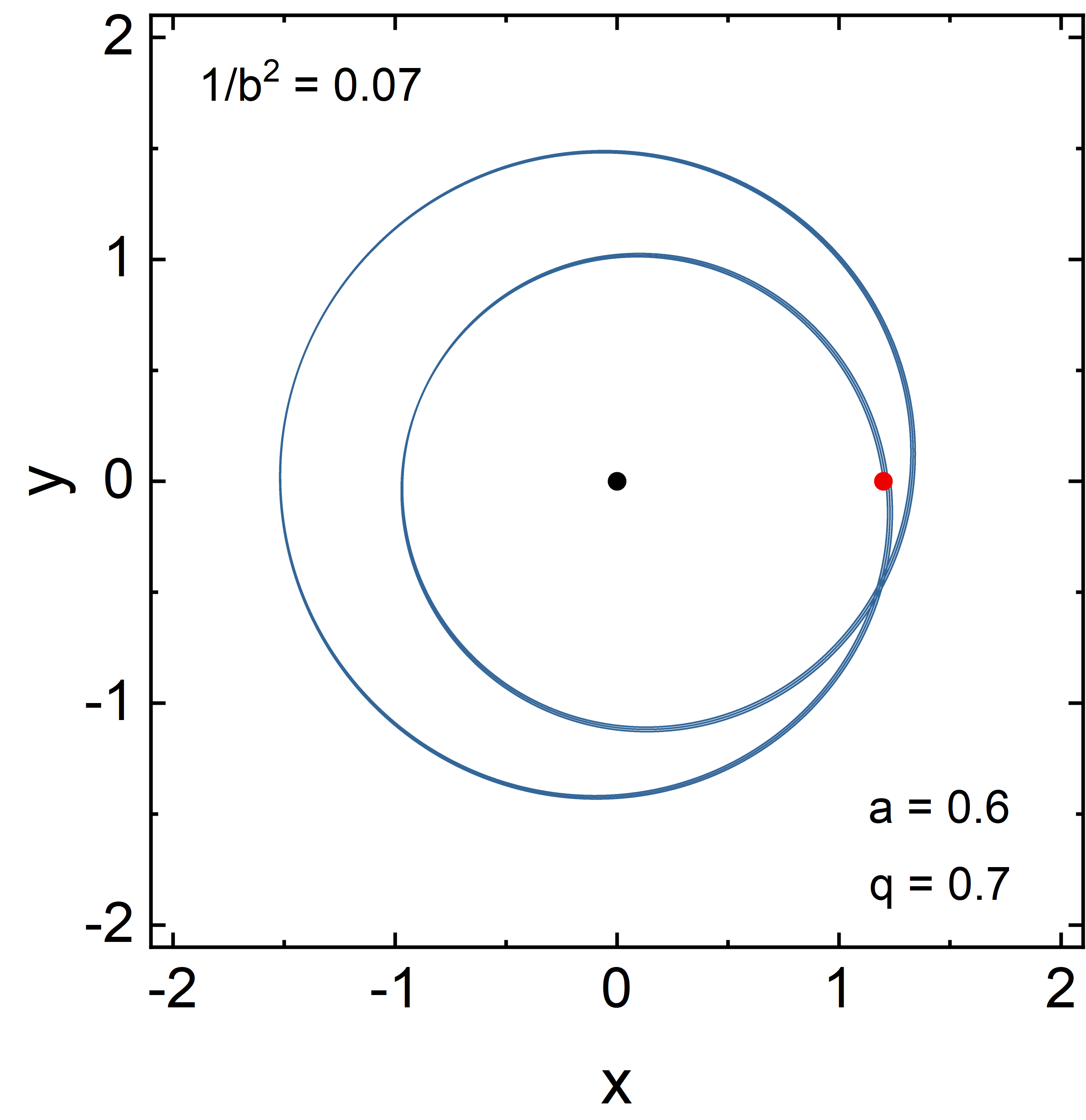}}
\subfigure{
\includegraphics[width=3.8cm,height=3.8cm]{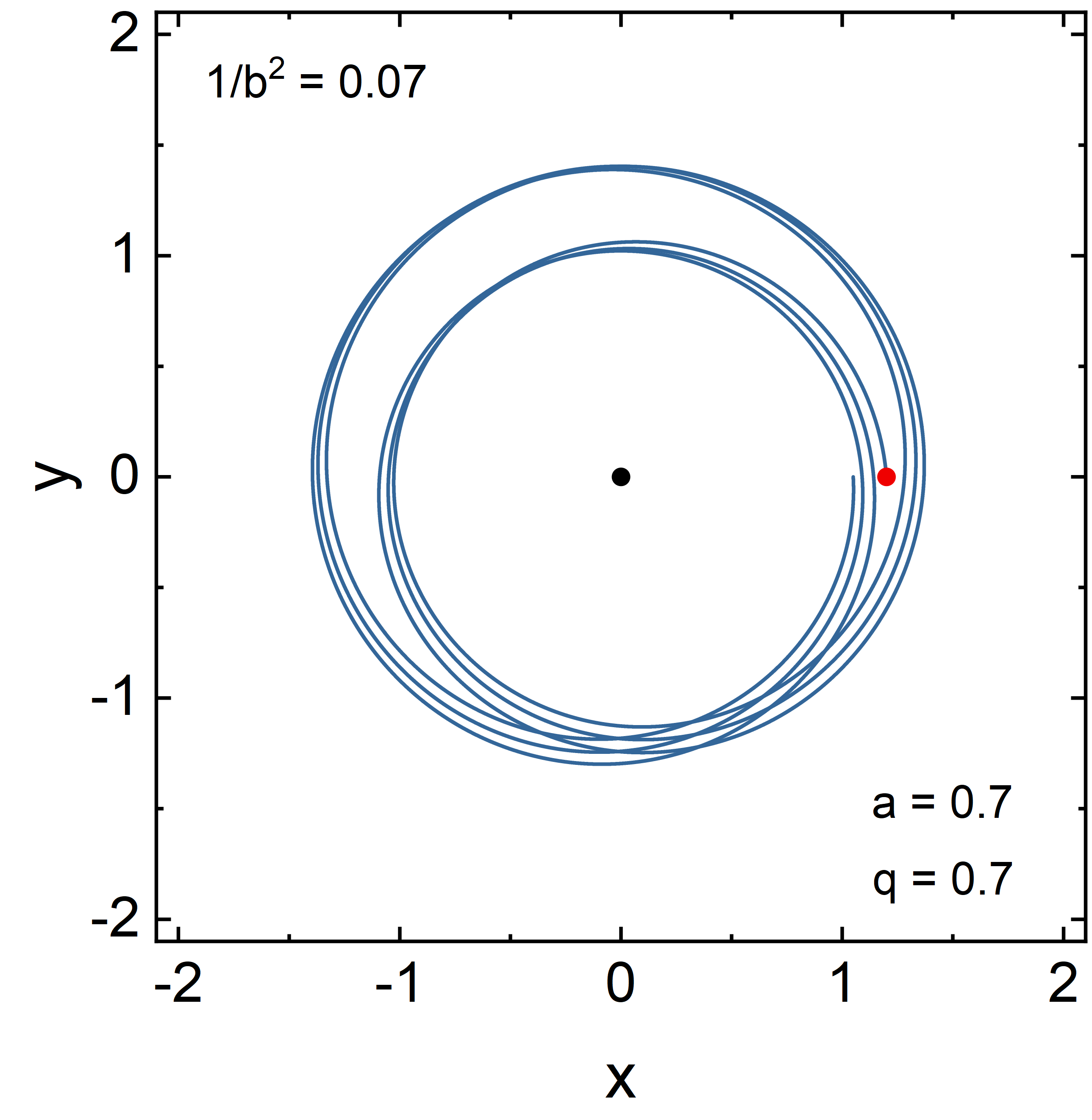}}\\
\subfigure{
\includegraphics[width=3.8cm,height=3.8cm]{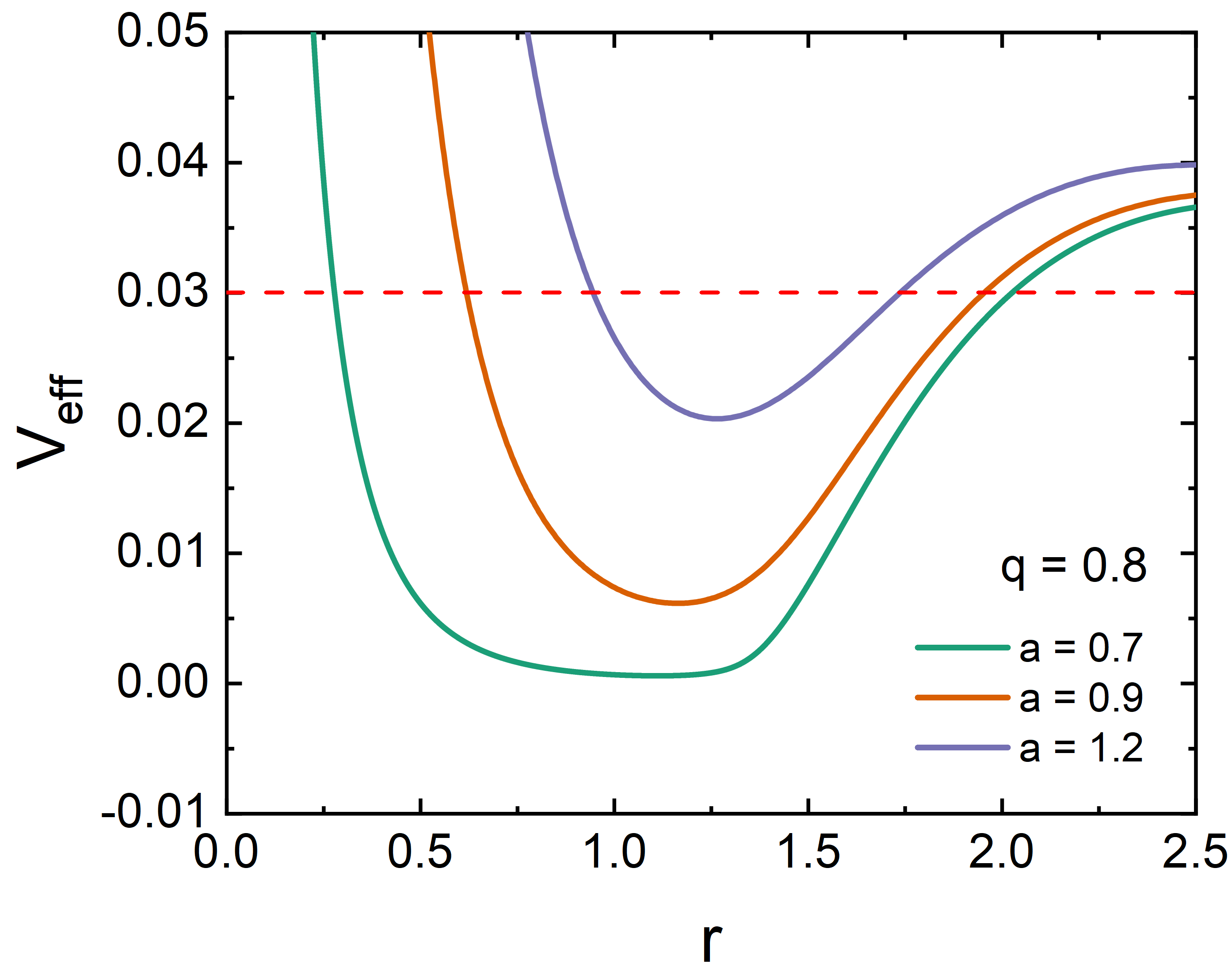}}
\subfigure{
\includegraphics[width=3.8cm,height=3.8cm]{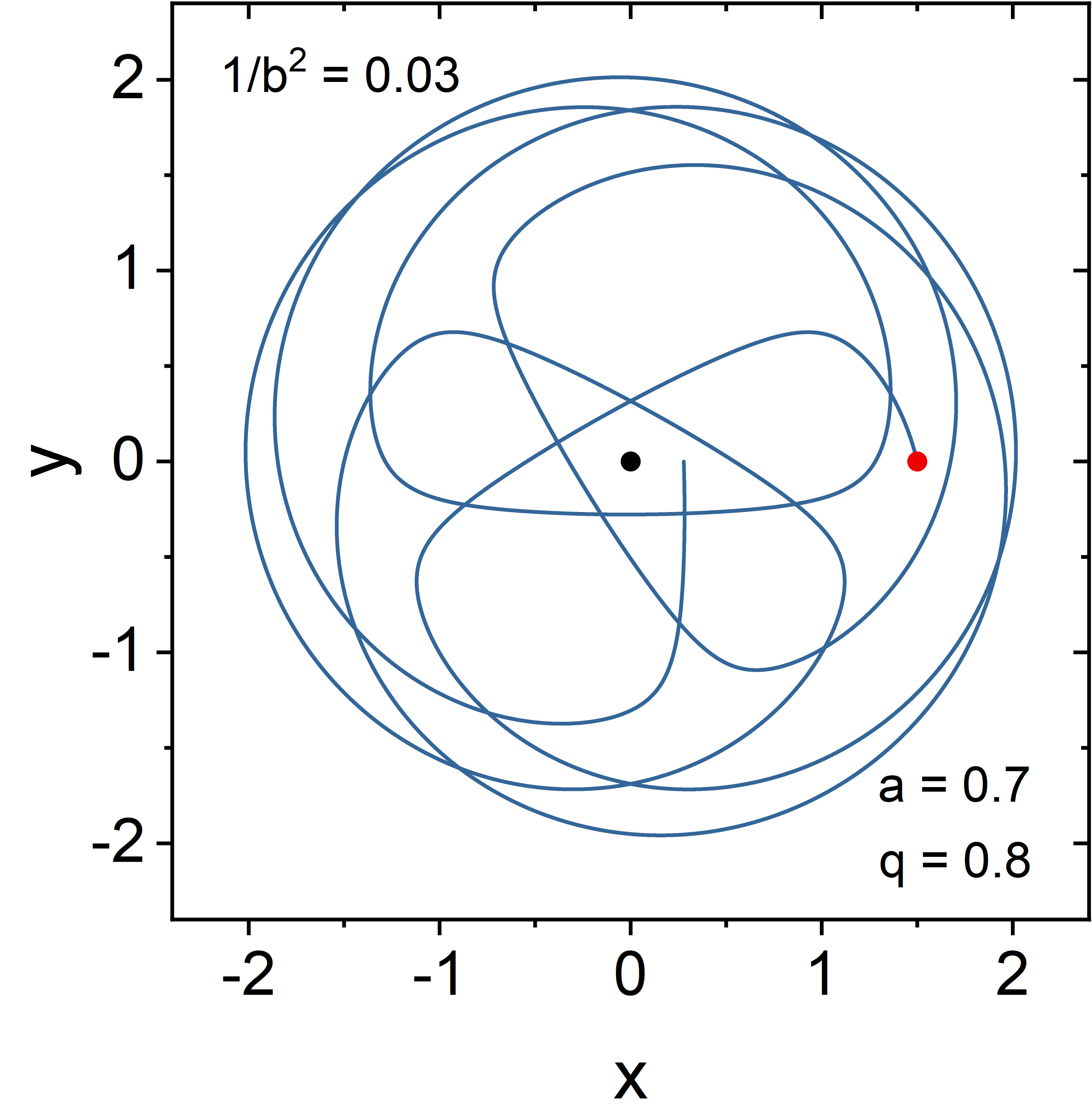}}
\subfigure{
\includegraphics[width=3.8cm,height=3.8cm]{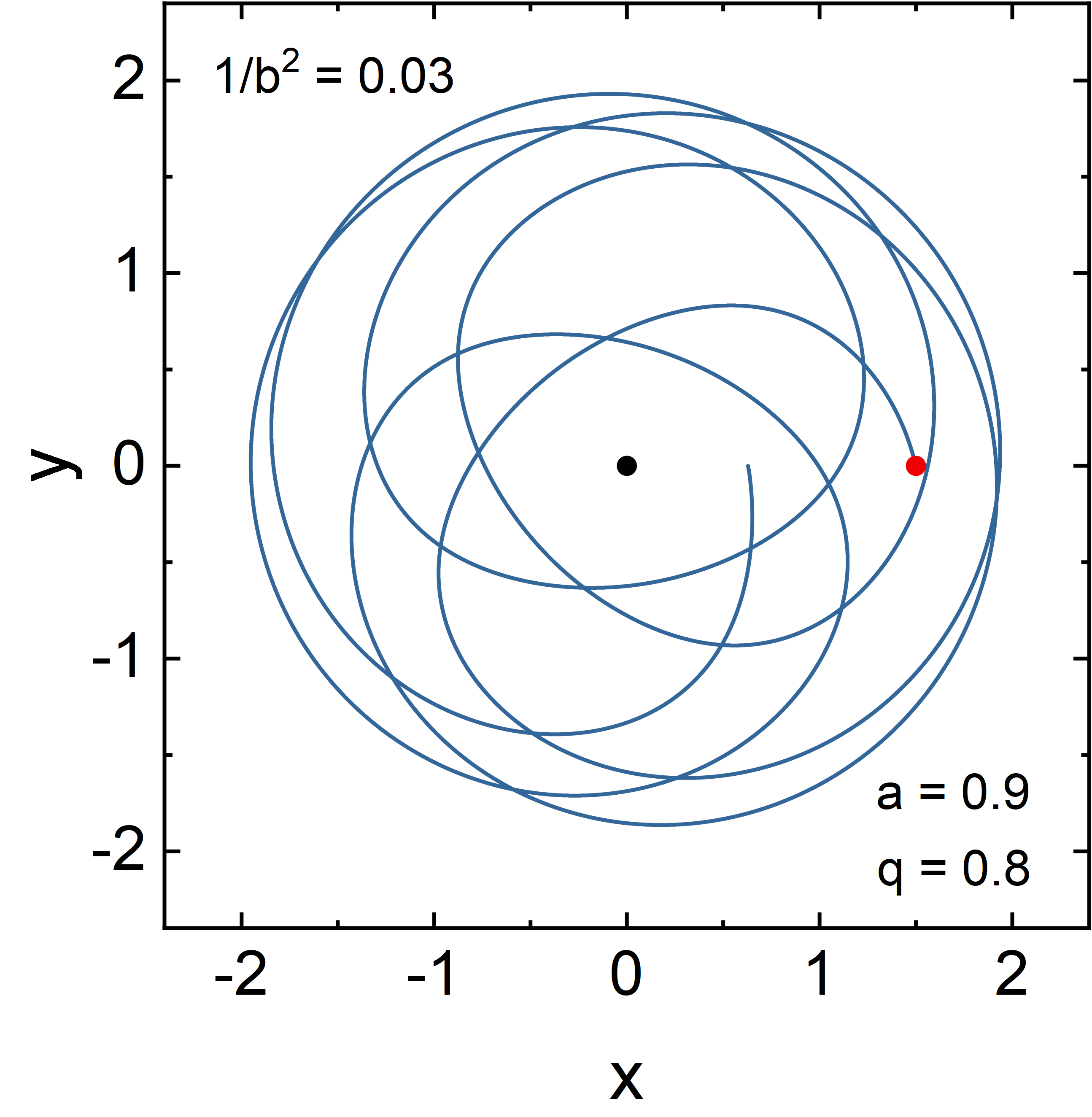}}
\subfigure{
\includegraphics[width=3.8cm,height=3.8cm]{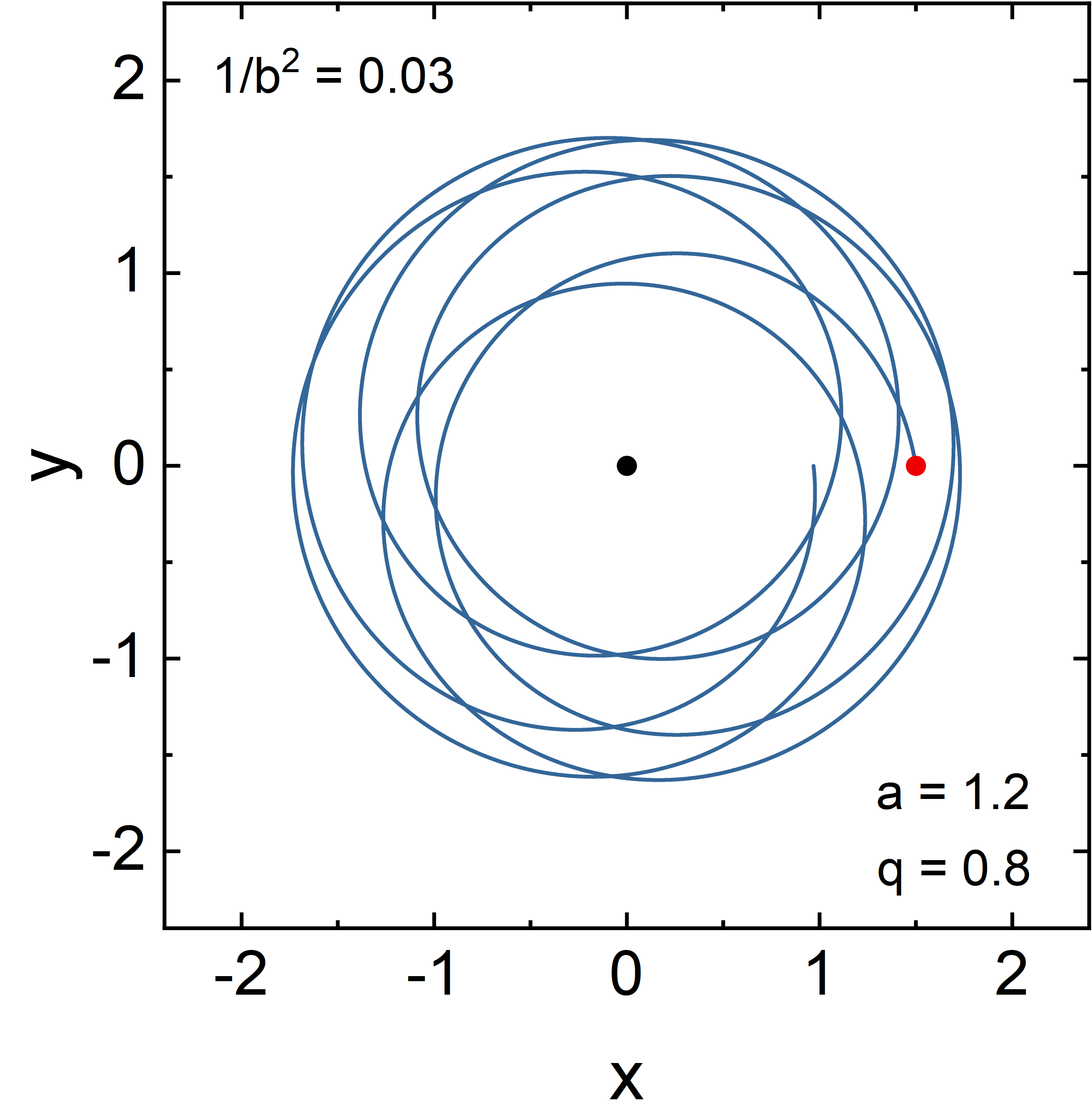}}	
\end{center}
		\caption{The shapes of the effective potential and bound orbits for $q=0.7$ (top) and $q=0.8$ (bottom).}
		\label{fig:BOUND}		
		\end{figure}
\begin{figure}[!htbp]
		\begin{center}
		\subfigure{ 
			\includegraphics[height=.28\textheight,width=.33\textheight, angle =0]{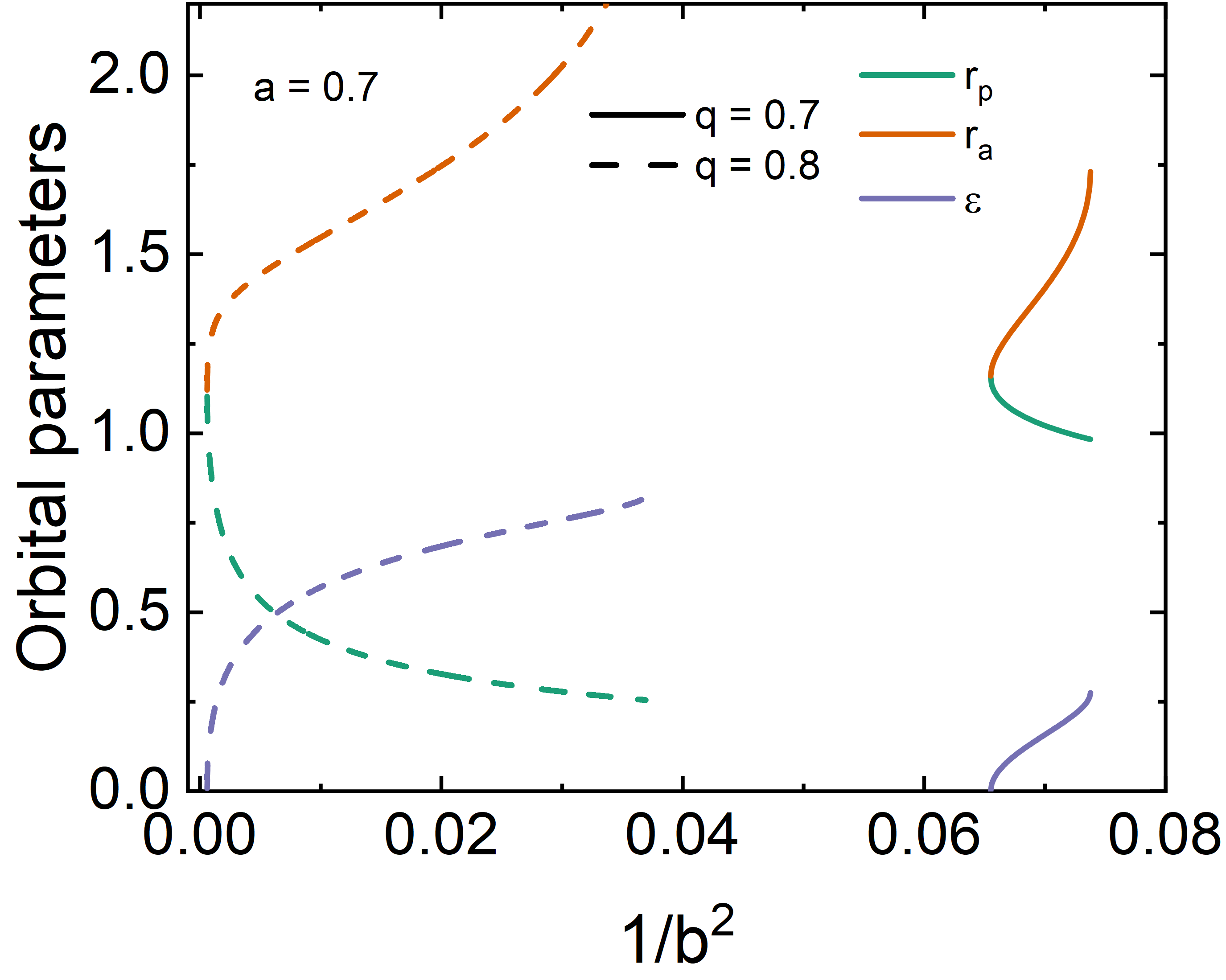}
			\label{fig:eccentricityb}
		}	 
  		\subfigure{  
			\includegraphics[height=.28\textheight,width=.33\textheight, angle =0]{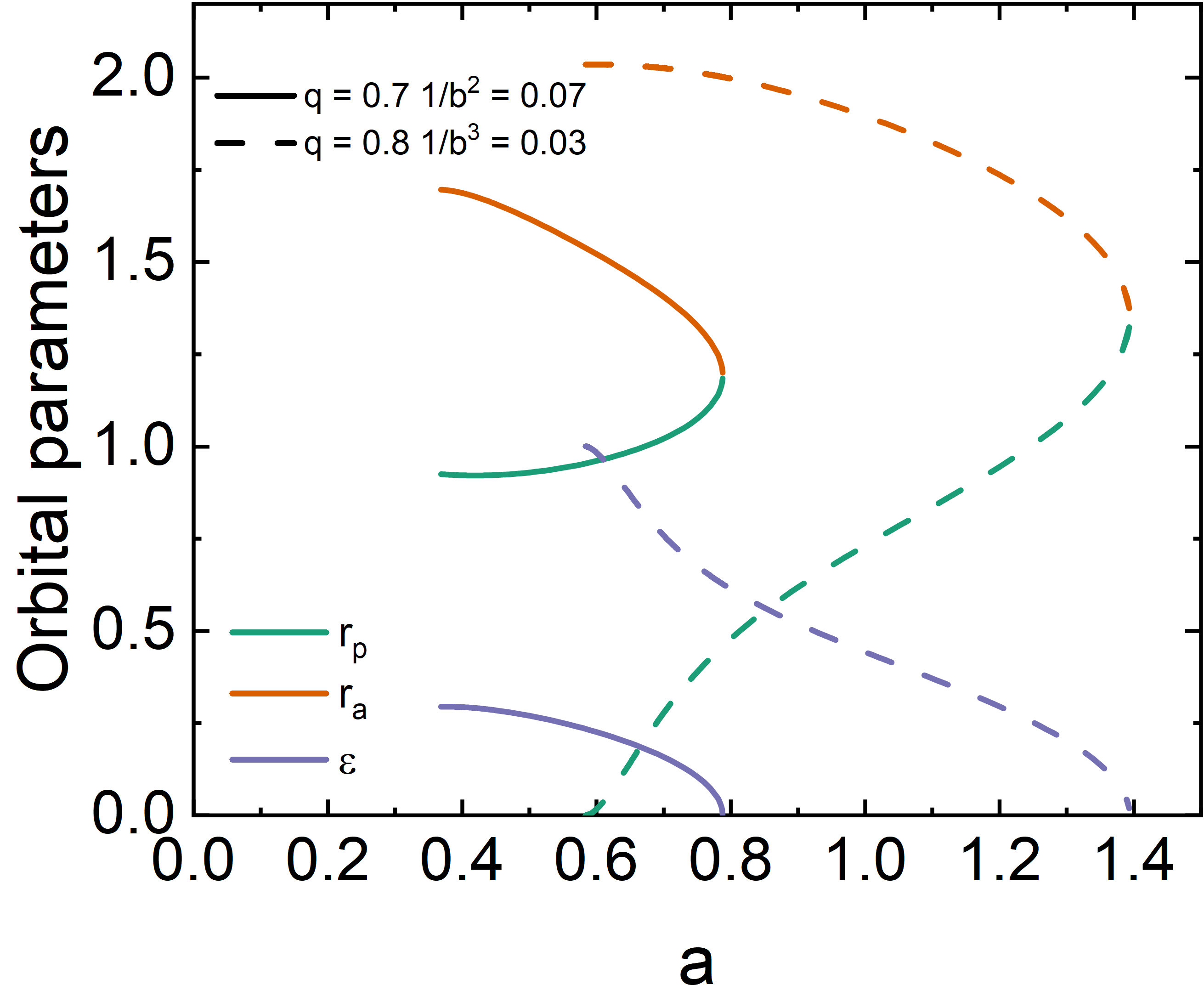}
			\label{fig:eccentricity}
		}	 
  		\end{center}	
		\caption{The orbital parameters as a function of the inverse square of the impact parameter $1/b^2$ (left) and coupling parameter $a$ (right).}
	\label{fig:matter}	
		\end{figure}

In addition to these unbound orbits, bound orbits can also be admitted when the radial effective potential of the solution exhibits an extremum, corresponding to the presence of LRs. The bound orbits for $q<q_{c}$ and $q\geq q_{c}$ are shown in Fig.~\ref{fig:BOUND}. All these orbits move from one apastron $r_a$ to periastron $r_p$ and then to the next apastron, where $r_a$ and $r_p$ can be given by the roots of equation $V_{eff}=1/b^2$ (see the leftmost panel). The apastron and periastron can define the orbital eccentricity $\epsilon=\frac{r_p-r_a}{r_p+r_a}$, which is an important parameter describing the shape of orbits. Fig.~\ref{fig:eccentricity} reveals that the eccentricity grows with increasing $1/b^2$ but decreases with increasing $a$.  
    \section{CONCLUSION}\label{sec: conclusion}
    In this paper, by means of numerical methods, we study the scalarization of the Bardeen spacetime throughout the whole spacetime and analyze the trajectory of photons around this model. Our results reveal that SBS solutions emerge once the scalarization coupling parameter $a$ surpasses the threshold $a_t$, with the scalar charge $Q_s$ increasing from zero initially and subsequently decreasing as $a$ grows.

    These SBS solution exhibit the following features. First, their ADM masses are always smaller than those of the corresponding pure Bardeen solutions with the same parameters. In other words, the introduction of the scalar field reduces the mass. Second, when $q$ is relatively large, solutions with LRs can emerge in certain regions of the domain of existence of the solution. The eccentricity of the bound orbits associated with these solutions  increases with increasing the inverse square of the impact parameter $1/b^2$, while it decreases with scalarization parameter $a$. Moreover, in the limit $a \to \infty$, our results indicate that the spacetime of these SBSs is very likely to approach a flat space.

    However, SBS can be classified into two distinct solution families depending on whether the magnetic charge $q$ is smaller than or larger than the critical charge $q_c$. For $q<q_c$, as $a$ approaches $a_t$, the real scalar field nearly vanishes, and the SBS almost reduces to the pure Bardeen solution. This corresponds to the general ``spontaneous" scalarization. In contrast, it is interesting that for $q \geq q_c$, when $a$ approaches $a_t$, the SBS transitions into a frozen and the scalar field can no longer be regarded as a small perturbation. In the background of frozen SBS, trajectories of photon experience extremely strong deflections near the critical horizon $r_{cH}$ and can pass very close to the origin. Moreover, inside and particularly near $r_{cH}$, photons take an extremely long time in motion from the perspective of an observer at infinity. Nevertheless, frozen SBSs remain clearly distinct from BHs: their null geodesics are still complete, allowing test photons to traverse the entire spacetime.
    
    Moreover, it is worthwhile to point that for these two types of SBSs, although their scalar charges approach zero in the limit $a \to a_t$, the underlying mechanisms responsible for this phenomenon are distinct. For $q < q_c$, the scalar charge vanishes due to the fact that the scalar field nearly vanishes entirely throughout the spacetime. In contrast, for $q \geq q_c$, the scalar charge approaches zero because the scalar field is almost distributed near the horizon, with its value approaching nearly zero outside the horizon.

    In Ref.~\cite{Zhang:2024bfu}, the authors solved the model only outside the horizon. Under this restriction, hairy BHs can appear, and near the scalarization threshold, the solution reduces to the pure Bardeen BH. In contrast, in our work, when the model is solved over the entire spacetime, in the limit $a\rightarrow a_t$  from the left, the solution of equations of motion ``discontinuously" transitions from a pure Bardeen black hole to a frozen SBS configuration and only hairy frozen solutions are found. For BH solutions, they only exist in the regime $a < a_t$, where the scalar field vanishes. In other words, when full spacetime is considered, hairy Bardeen BH solutions do not seem to emerge. Therefore, in this sense, the result of our work suggests a ``no-hair" theorem from the ``whole space-time perspective".

    Several extensions of our study merit exploration. First, in this article, we only considered the spontaneous scalarization of the Bardeen model. However, in the Standard Model, many fundamental particles are represented by spinors and vectors. This naturally raises an intriguing question: What would happen if the scalar field were replaced by a vector field or a spinor? Secondly, in addition to the photon orbit around SBSs, the accretion disk composed of timelike particles has an influence on gravitational wave signals and will probably reveal more observational characteristics. We intend to investigate this aspect in our future work. Finally, the stability of SBS also remains an important issue for our future investigation.
    \section*{ACKNOWLEDGEMENTS}
	This work is supported by the National Natural Science Foundation of China (Grant No.~12275110 and No.~12247101) and the National Key Research and Development Program of China (Grant No.~2022YFC2204101 and 2020YFC2201503).).

\end{document}